\documentclass[preprintnumbers,amsmath,amssymb,showpacs,twocolumn]{revtex4}

\usepackage{graphicx}
\usepackage{dcolumn}
\usepackage{bm}

\begin{document}

\title{Optimal Controlled teleportation via several kinds of three-qubit states}

\author{Ting Gao$^{1,2}$, Feng-Li Yan$^{2,3}$, and You-Cheng Li$^3$}

\affiliation {$^1$ College of Mathematics and Information Science, Hebei Normal University, Shijiazhuang 050016, China\\
$^2$ Max-Planck-Institut f\"{u}r Quantenoptik, Hans-Kopfermann-Str. 1, D-85748 Garching, Germany \\
$^3$ College of Physics and Information Engineering, Hebei Normal University, Shijiazhuang 050016, China}

\date{\today}

\begin{abstract}
The  probability of successfully controlled teleportating an unknown qubit using a general three-particle state
is investigated. We give the analytic expressions of  maximal probabilities of successfully controlled
teleportating an unknown qubit via several kinds of tripartite states including a tripartite GHZ state and a
tripartite W-state. \end{abstract}

\pacs{ 03.67.Hk, 89.70.+c}

\maketitle

\section{Introduction}

 Bennett \emph{et al.} \cite{BBCJPW} showed that an arbitrary unknown state of a
qubit could be teleported from a sender to a spatially distant receiver  with the aid of long-range
Einstein-Podolsky-Rosen (EPR) correlations and the transmission of two bits of classical information. Since
then, quantum teleportation has been developed by many authors \cite{FujiiPRA2003, BaAnPRA2003, SLKPpra2001,
 BTBSRBSL, JBS, VaidmanPRA1994, BKprl1998,  BDEFMS, GRpra, GYWQIC,KB,YCHpra2004, PatiPRA2000, APpla, PAjob, DLLZWpra2005,YW}
  due to its important applications in quantum communication \cite{YZ-EPJB2004}
 and quantum computation. In past several years quantum teleportation has been also experimentally
 demonstrated by several groups \cite{BPMEWZ,NKL}.

  The controlled quantum
 teleportation scheme  was  presented by Karlsson and Bourennane \cite{KB}.  In the scheme, an unknown state can be
  perfectly
transported from one place to another place via previously shared GHZ state by means of local operations and
classical communications (LOCC) under the permission of the third party. The signal state can not be transmitted
unless all three sides agree to cooperate. The controlled quantum teleportation is useful in networked quantum
information processing and cryptographic conferencing \cite{AT, BHMpra1996, Townsend, BVKpra2003}, and
controlled quantum secure direct communication \cite{Gaozna} and has other interesting applications, such as in
opening account on the agreement of managers in a network. Recently, a number of works on controlled quantum
teleportation have also been proposed \cite{YCHpra2004, PatiPRA2000, DLLZWpra2005,YW}, where they restrict
themselves to the special quantum channels, such as GHZ state or W state. If a nonmaximally entangled state is
taken as quantum channel, then one can not teleport a qubit with unit probability  and unit fidelity. However,
it is possible to teleport a qubit with a probability $p<1$, which is called probabilistic quantum teleportation
\cite{APpla,PAjob}. More recently, the probabilistic scheme has been generalized to teleport $N$ qubits
\cite{GRpra,GYWQIC}.

The entanglement property lies at the very heart of quantum information theory. The reason is that entanglement
is the physical resource to perform some of the most important quantum information tasks, such as quantum
teleportation, quantum computation etc. In \cite{VPCirac-prl(04)}, Verstraete, Popp, and Cirac introduced a new
concept which they called localizable entanglement (LE). This quantity not only has a very well defined physical
meaning that treats entanglement as a truly physical resource, but also establishes a very close connection
between entanglement and correlation functions. The LE  $E_{ij}$ is defined as the maximum of the average
entanglement between the spins $i$ and $j$ over all possible outcomes
\begin{equation}\label{}
E_{ij}=\max_\varepsilon \sum_s p_s E(|\phi_s\rangle),
\end{equation}
where $p_s$ denotes the probability to obtain the two-spin state $|\phi_s\rangle$ after performing the
measurement $|s\rangle$ in the rest of the system,  $E(|\phi_s\rangle)$ is the chosen measure of entanglement of
$|\phi_s\rangle$.
 The determination of the LE is a formidable task since it involves optimization over all possible
local measurement strategies, and thus can not be determined in general. However,  Verstaete, Popp, and Cirac
gave  tight upper bound and lower pound in case of  $E(|\phi_s\rangle)$ being the concurrence of
$|\phi_s\rangle$. We determined  the exact value of this kind of LE of the general tripartite state, and
obtained the analytic expression of another kind of LE, the maximal successful probability of controlled
teleporting a qubit of unknown information from a sender to a remote receiver via the control of a third agent
by the use of a general three-qubit state \cite{GYL1}.

In this paper, we give the exact values of the  maximal probabilities of successfully controlled teleportating
an unknown qubit via many kinds of tripartite states including a tripartite GHZ state and a tripartite W-state.

The paper is outlined as follows. In  Section II, we present a scheme for controlled probabilistic quantum
teleportation of
 an arbitrary unknown qubit with a general three-qubit state. Furthermore, the successful probability of this
  teleportation is also obtained.  In Section III,  we show how to select measuring basis to reach the maximal successful
  probability of controlled teleporting the complete information about an arbitrary
unknown state of a qubit using some kinds of three-qubit states.
 A brief  summary is given in Section IV.

\section{The controlled quantum teleportation using a general three-particle state}

Ac\'{i}n  \emph{et al.} \cite{AACJLT} gave the minimal decomposition of any pure three-qubit state in terms of
orthogonal product states built from local bases------a generalization of the two-quantum-bit  Schmidt
decomposition. They  proved that for any pure three-quantum-bit state the existence of local bases which allow
one to build a set of five orthogonal product states in terms of which the state can be written in a unique
form. That is, for every pure state of a composite system, 123, there exist orthonormal states $|0\rangle_1$,
$|1\rangle_1$ for system 1, orthonormal states $|0\rangle_2$, $|1\rangle_2$ for system 2, and orthonormal states
$|0\rangle_3$, $|1\rangle_3$ for system 3  such  that
\begin{equation}\label{quantumchannel}
\begin{array}{ll}
|\Psi\rangle_{123}= & a_0|000\rangle_{123}+a_1e^{\texttt{i}\mu}|100\rangle_{123}+a_2|101\rangle_{123} \\
  & +a_3|110\rangle_{123}+a_4|111\rangle_{123}, \\
 &  a_i\geq 0, ~ 0\leq
  \mu\leq \pi, ~ \Sigma_{i=0}^{4} a_i^2=1. \\
\end{array}
\end{equation}
It is uniquely characterized by the five entanglement parameters.

Suppose that Alice is to deliver an unknown state to a distant receiver Bob supervised by the controller Charlie
via a quantum channel of a normalized general pure three-qubit
 state in (\ref{quantumchannel}),
 where particle 1 belongs to Charlie, particle 2 is in Alice's
 side, while Bob has particle 3. Let  $a_0\neq 0$ through out the paper. Since if $a_0=0$, then $|\Psi\rangle_{123}$ is a tensor product state of a pure state of
 particle 1 and a pure state of particles 2 and 3, but not a true tripartite entangled state.
Bob can get the qubit of quantum information carried by the unknown state only if he obtains the permission of
Charlie (i.e., Charlie is trustworthy and cooperative).

After  getting the approval of Charlie, Alice and Bob begin their teleportation under the control of Charlie.

The controller Charlie  measures his particle in the basis
 \begin{equation}\label{measurebasis}
 \begin{array}{c}
  |x\rangle=\cos\frac{\theta}{2}|0\rangle+e^{\texttt{i}\varphi}\sin\frac{\theta}{2}|1\rangle, \\
  |x\rangle^\perp=\sin\frac{\theta}{2}|0\rangle-e^{\texttt{i}\varphi}\cos\frac{\theta}{2}|1\rangle, \\
 \end{array}
\end{equation}
and broadcasts his measurement result. Here $ \theta\in[0,\pi], \varphi\in[0,2\pi]$.

The tripartite state $|\Psi\rangle_{123}$ can be reexpressed as
\begin{equation}\label{quantumchanneldecomposition}
\begin{array}{lll}
|\Psi\rangle_{123} &=& \sqrt{p_1}|x\rangle_1|\Phi_1\rangle_{23}+\sqrt{p_2}|x\rangle_1^\perp|\Phi_2\rangle_{23}.
\end{array}
\end{equation}
Here
\begin{eqnarray}
 p_1  &=& \sin^2\frac{\theta}{2}+a_0^2\cos\theta+a_0a_1\cos(\mu-\varphi)\sin\theta, \label{p1}\\
  p_2 &=& \cos^2\frac{\theta}{2}-a_0^2\cos\theta-a_0a_1\cos(\mu-\varphi)\sin\theta, \label{p2}\\
 |\Phi_1\rangle_{23} &=& \frac{1}{\sqrt{p_1}}[(a_0\cos\frac{\theta}{2}+a_1e^{\texttt{i}(\mu-\varphi)}\sin\frac{\theta}{2})|00\rangle_{23} \nonumber\\
&&   +a_2e^{-\texttt{i}\varphi}\sin\frac{\theta}{2}|01\rangle_{23}+a_3e^{-\texttt{i}\varphi}\sin\frac{\theta}{2}|10\rangle_{23} \nonumber\\
&&  +a_4e^{-\texttt{i}\varphi}\sin\frac{\theta}{2}|11\rangle_{23}], \\
  |\Phi_2\rangle_{23} &=& \frac{1}{\sqrt{p_2}}[(a_0\sin\frac{\theta}{2}-a_1e^{\texttt{i}(\mu-\varphi)}\cos\frac{\theta}{2})|00\rangle_{23} \nonumber\\
&& -a_2e^{-\texttt{i}\varphi}\cos\frac{\theta}{2}|01\rangle_{23}-a_3e^{-\texttt{i}\varphi}\cos\frac{\theta}{2}|10\rangle_{23} \nonumber\\
&& -a_4e^{-\texttt{i}\varphi}\cos\frac{\theta}{2}|11\rangle_{23}].
\end{eqnarray}

After Charlie's measurement, the quantum channel is collapsed to $|\Phi_1\rangle_{23}$ and $|\Phi_2\rangle_{23}$
with probability $p_1$ and $p_2$, respectively.

 By Schmidt decomposition,
\begin{eqnarray}\label{schmidtdecomposition}
|\Phi_1\rangle_{23}&=& \sqrt{\lambda_{10}}|0_2'0_3'\rangle+\sqrt{\lambda_{11}}|1'_21'_3\rangle,\\
|\Phi_2\rangle_{23}&=& \sqrt{\lambda_{20}}|\bar{0}_2\bar{0}_3\rangle+\sqrt{\lambda_{21}}|\bar{1}_2\bar{1}_3\rangle,
\label{schmidtdecomposition1}
\end{eqnarray}
where $\{0_2',1_2'\}$ and  $\{\bar{0}_2,\bar{1}_2\}$ ( $\{0_3',1_3'\}$, and  $\{\bar{0}_3,\bar{1}_3\}$ ) are orthonormal bases of system 2
(system 3), and Schmidt coefficients
\begin{equation}\label{}
\begin{array}{cc}
  \lambda_{10}=\frac{1-\sqrt{1-C_1^2}}{2}, &  \lambda_{11}= \frac{1+\sqrt{1-C_1^2}}{2}, \\
  \lambda_{20}= \frac{1-\sqrt{1-C_2^2}}{2}, &  \lambda_{21} = \frac{1+\sqrt{1-C_2^2}}{2}.
\end{array}
\end{equation}
Here
 $C_1 = \frac{|a_0a_4e^{-\texttt{i}\varphi}\sin\theta+2(a_1a_4e^{\texttt{i}\mu}
-a_2a_3)e^{-2i\varphi}\sin^2\frac{\theta}{2}|}{p_1}$ and $ C_2=
\frac{|a_0a_4e^{-\texttt{i}\varphi}\sin\theta-2(a_1a_4e^{\texttt{i}\mu}
-a_2a_3)e^{-2i\varphi}\cos^2\frac{\theta}{2}|}{p_2}$ are the concurrence of $|\Phi_1\rangle_{23}$  and
$|\Phi_2\rangle_{23}$, respectively.

 For simplicity, we write (\ref{schmidtdecomposition}) and (\ref{schmidtdecomposition1}) as
\begin{eqnarray}\label{simpleschmidtdecomposition}
|\Phi_1\rangle_{23}&=& \sqrt{\lambda_{10}}|0_20_3\rangle+\sqrt{\lambda_{11}}|1_21_3\rangle,\\
|\Phi_2\rangle_{23}&=& \sqrt{\lambda_{20}}|0_20_3\rangle+\sqrt{\lambda_{21}}|1_21_3\rangle.
\end{eqnarray}

Suppose that the unknown quantum state the sender Alice wants to teleport to  Bob is
\begin{equation}\label{unknownstate}
    |\psi\rangle_4=\alpha|0\rangle_4+\beta|1\rangle_4, |\alpha|^2+|\beta|^2=1.
\end{equation}

If the measurement outcome of Charlie is $|x\rangle_1$, then the collect state of particles 2, 3 and 4 is
\begin{equation}\label{Phi_1}
\begin{array}{ll}
&  |\psi\rangle_4|\Phi_1\rangle_{23} \\
 = &
  \sqrt{\lambda_{10}|\alpha|^2+\lambda_{11}|\beta|^2}[\frac{1}{\sqrt{2}}(|\phi^+\rangle_{24}\frac{\alpha\sqrt{\lambda_{10}}|0\rangle_3
  +\beta\sqrt{\lambda_{11}}|1\rangle_3}{\sqrt{\lambda_{10}|\alpha|^2+\lambda_{11}|\beta|^2}} \\
&  +|\phi^-\rangle_{24}\frac{\alpha\sqrt{\lambda_{10}}|0\rangle_3
  -\beta\sqrt{\lambda_{11}}|1\rangle_3}{\sqrt{\lambda_{10}|\alpha|^2+\lambda_{11}|\beta|^2}})] \\
  & +\sqrt{\lambda_{11}|\alpha|^2+\lambda_{10}|\beta|^2}[\frac{1}{\sqrt{2}}(|\psi^+\rangle_{24}\frac{\beta\sqrt{\lambda_{10}}|0\rangle_3
  +\alpha\sqrt{\lambda_{11}}|1\rangle_3}{\sqrt{\lambda_{11}|\alpha|^2+\lambda_{10}|\beta|^2}}  \\
&  +|\psi^-\rangle_{24}\frac{\beta\sqrt{\lambda_{10}}|0\rangle_3
  -\alpha\sqrt{\lambda_{11}}|1\rangle_3}{\sqrt{\lambda_{11}|\alpha|^2+\lambda_{10}|\beta|^2}})],
\end{array}
\end{equation}
where $|\phi^\pm\rangle=\frac {1}{\sqrt 2}(|00\rangle\pm|11\rangle), |\psi^\pm\rangle=\frac {1}{\sqrt
2}(|01\rangle\pm|10\rangle)$. Alice makes a Bell measurement on her particles 2 and 4. She obtains
$|\phi^+\rangle_{24}$, $|\phi^-\rangle_{24}$, $|\psi^+\rangle_{24}$, and $|\psi^-\rangle_{24}$ with probability
$\left(\frac{\sqrt{\lambda_{10}|\alpha|^2+\lambda_{11}|\beta|^2}}{\sqrt{2}}\right)^2$,
$\left(\frac{\sqrt{\lambda_{10}|\alpha|^2+\lambda_{11}|\beta|^2}}{\sqrt{2}}\right)^2$,
$\left(\frac{\sqrt{\lambda_{11}|\alpha|^2+\lambda_{10}|\beta|^2}}{\sqrt{2}}\right)^2$, and
$\left(\frac{\sqrt{\lambda_{11}|\alpha|^2+\lambda_{10}|\beta|^2}}{\sqrt{2}}\right)^2$, respectively. Then she
conveys  her measurement outcome to Bob   over a classical communication channel.

In order to achieve teleportation, Bob needs to introduce an auxiliary particle $b$ with the initial state $|0\rangle_b$ and performs a
collective unitary
transformation $U_{3b}=\left(%
\begin{array}{cccc}
  1 & 0 & 0 & 0 \\
  0 & 1 & 0 & 0 \\
  0 & 0 & \frac{\sqrt{\lambda_{10}}}{\sqrt{\lambda_{11}}} & \sqrt{1-\frac{\lambda_{10}}{\lambda_{11}}} \\
  0 & 0 & -\sqrt{1-\frac{\lambda_{10}}{\lambda_{11}}} & \frac{\sqrt{\lambda_{10}}}{\sqrt{\lambda_{11}}} \\
\end{array}%
\right)$ on the state of particles 3 and $b$. Then the measurement on his auxiliary particle $b$ follows. If his measurement result is
$|0\rangle_b$, Bob can fix up the state of his particle 3, recovering $|\psi\rangle$, by applying an appropriate local unitary operation.   The
achievable successful probability of teleporting the unknown state in (\ref{unknownstate}) via $|\Phi_1\rangle_{23}$ is
$2\left(\frac{\sqrt{\lambda_{10}|\alpha|^2+\lambda_{11}|\beta|^2}}{\sqrt{2}}\right)^2\left(\frac{\sqrt{\lambda_{10}}}{\sqrt{\lambda_{10}|\alpha|^2+\lambda_{11}|\beta|^2}}\right)^2
+2\left(\frac{\sqrt{\lambda_{11}|\alpha|^2+\lambda_{10}|\beta|^2}}{\sqrt{2}}\right)^2\left(\frac{\sqrt{\lambda_{10}}}{\sqrt{\lambda_{11}|\alpha|^2+\lambda_{10}|\beta|^2}}\right)^2
=2\lambda_{10}$. Similarly, if the measurement result of Charlie is $|x\rangle^\perp_1$, the achievable successful probability of teleporting
the state in (\ref{unknownstate}) via $|\Phi_2\rangle_{23}$ is $2\lambda_{20}$.

Therefore, probability $p$ of successfully controlled teleporting an unknown qubit (\ref{unknownstate}) using a general three-particle state in
(\ref{quantumchannel}) is
\begin{equation}\label{probability}
 \begin{array}{rl}
   p= & 2p_1\lambda_{10}+2p_2\lambda_{20} \\
= & 1-\left(\sqrt{P(\theta,\varphi)} + \sqrt{Q(\theta,\varphi)}\right).
 \end{array}
\end{equation}
Here {\small
\begin{equation}\label{p}
\begin{array}{ll}
& P(\theta, \varphi)=p_1^2(1-C_1^2) \\
 =&\frac{1}{4}a_0^2a_1^2\cos2(\varphi -\mu) +
  3{a_1}{a_2}{a_3}{a_4}\cos\mu + \frac{1}{8}(3 - 4a_0^2 \\
& + 4a_0^4 + 2a_0^2a_1^2 - 12a_2^2a_3^2 -
     4a_0^2a_4^2 - 12a_1^2a_4^2)\\
 &  +  \frac{1}{8}\cos2\theta( 1 - 4a_0^2 + 4a_0^4 - 2a_0^2a_1^2 -
  2a_0^2a_1^2\cos2(\varphi -\mu) \\
&  - 4a_2^2a_3^2 + 8{a_1}{a_2}{a_3}{a_4}\cos\mu +
       4a_0^2a_4^2 - 4a_1^2a_4^2)\\
& -\cos\theta(\frac{1}{2} - a_0^2 - 2a_2^2a_3^2 +4{a_1}{a_2}{a_3}{a_4}\cos\mu-
     2a_1^2a_4^2) \\
&  + {a_0}\sin\theta[ 2{a_2}{a_3}{a_4}\cos\varphi +
     {a_1}( 1 - 2a_4^2)\cos(\varphi -\mu)]\\
&  - \frac{1 }{2}{a_0}\sin2\theta[ 2{a_2}{a_3}{a_4}\cos\varphi +
      {a_1}(1 - 2a_0^2 - 2a_4^2) \cos (\varphi -\mu) ],
  \end{array}
\end{equation}
\begin{equation}\label{q}
  \begin{array}{ll}
& Q(\theta,\varphi)=p_1^2(1-C_2^2) \\
=& \frac{1}{4}a_0^2a_1^2\cos2(\varphi -\mu) +
  3{a_1}{a_2}{a_3}{a_4}\cos\mu + \frac{1}{8}(3 - 4a_0^2  \\
&   + 4a_0^4 + 2a_0^2a_1^2 - 12a_2^2a_3^2 -
     4a_0^2a_4^2 - 12a_1^2a_4^2)\\
&  +  \frac{1}{8}\cos2\theta( 1 - 4a_0^2 + 4a_0^4 - 2a_0^2a_1^2 -
  2a_0^2a_1^2\cos2(\varphi -\mu) \\
& - 4a_2^2a_3^2 + 8{a_1}{a_2}{a_3}{a_4}\cos\mu +
       4a_0^2a_4^2 - 4a_1^2a_4^2)\\
& +\cos\theta(\frac{1}{2} - a_0^2 - 2a_2^2a_3^2 +4{a_1}{a_2}{a_3}{a_4}\cos\mu-
     2a_1^2a_4^2)  \\
&   - {a_0}\sin\theta [ 2{a_2}{a_3}{a_4}\cos\varphi +
     {a_1}( 1 - 2a_4^2)\cos(\varphi -\mu) ]\\
&  - \frac{1 }{2}{a_0}\sin2\theta( 2{a_2}{a_3}{a_4}\cos\varphi +
      {a_1}(1 - 2a_0^2 - 2a_4^2) \cos (\varphi -\mu) ).
  \end{array}
\end{equation}  }

Obviously,
\begin{equation}\label{q-p-relation}
\begin{array}{lllll}
  Q(\theta,\varphi) & = & P(\pi-\theta,\varphi+\pi) & \texttt{if} & \varphi\in[0,\pi], \\
  Q(\theta,\varphi) & = & P(\pi-\theta,\varphi-\pi) &  \texttt{if} & \varphi\in[\pi,2\pi],
\end{array}
 \end{equation}
 \begin{equation}\label{p-q-relation}
\begin{array}{lllll}
  P(\theta,\varphi) & = & Q(\pi-\theta,\varphi+\pi) & \texttt{if} & \varphi\in[0,\pi], \\
  P(\theta,\varphi) & = & Q(\pi-\theta,\varphi-\pi) &  \texttt{if} & \varphi\in[\pi,2\pi],
\end{array}
 \end{equation}
and
\begin{equation}\label{0-pi-2pi}
\begin{array}{lll}
 \sqrt{P(0,\varphi)} +\sqrt{Q(0,\varphi)} &=& \sqrt{P(\pi,\varphi)} +\sqrt{Q(\pi,\varphi)},  \\
 \sqrt{P(\theta,0)} +\sqrt{Q(\theta,0)} &=& \sqrt{P(\theta,2\pi)} +\sqrt{Q(\theta,2\pi)}.
\end{array}
\end{equation}

It is clear that the successful probability is the same as the above (\ref{probability})  if the operation order is changed. That is, if Alice
makes a Bell state measurement on her particles 2 and 4 first, the third party Charlie's measurement on his particle 1 follows, after that Bob
operates his particle to acquire a qubit of quantum information Alice sends, instead of the above operation order, then they can also achieve
the successful  probability (\ref{probability}) of controlled teleportation.

\section{The maximal successful probability of controlled quantum teleportation
using three-particle states with $a_1a_2a_3a_4\sin\mu =0$}

In \cite{GYL1}, we determined the analytic expression of the maximal successful probability of controlled
teleportation by using the general tripartite state (\ref{quantumchannel}) with $a_0a_1a_2a_3a_4\sin\mu\neq 0$.
 In this section, we give the analytic expression of the localizable
entanglement (LE), the maximum of probability of successfully controlled teleporting an unknown qubit state
(\ref{unknownstate}) via every three-qubit state (\ref{quantumchannel}) satisfying $a_1a_2a_3a_4\sin\mu =0$ and
investigate how to achieve it (that is, Charlie finds optimal measurement basis).

 Obviously, the maximum of
(\ref{probability}) is
\begin{equation}\label{max}
 \begin{array}{rl}
p_{\max}= &  \max\{p\} \\
 = & \max\{2p_1\lambda_{10}+2p_2\lambda_{20}\} \\
= & 1-\min\{\sqrt{P(\theta, \varphi)} + \sqrt{Q(\theta, \varphi)}\}.
 \end{array}
\end{equation}

In order to get the maximum $p_{\max}$ of $p$ in (\ref{max}), we need only  obtain the minimum of $\sqrt{P(\theta,\varphi)}
+\sqrt{Q(\theta,\varphi)}$
\begin{equation}\label{min23}
\begin{array}{ll}
   & \min\{\sqrt{P(\theta,\varphi)} +\sqrt{Q(\theta,\varphi)}\}. \\
 \end{array}
\end{equation}
In other words, to reach the maximal probability of exact controlled teleportation through an arbitrary
partially entangled quantum channel (\ref{quantumchannel}), the supervisor Charlie needs only to choose optimal
measurement basis, i.e. he selects $\theta_0$ and $\varphi_0$ such that $ \min\{\sqrt{P(\theta,\varphi)}
+\sqrt{Q(\theta,\varphi)}\}=\sqrt{P(\theta_0,\varphi_0)} +\sqrt{Q(\theta_0,\varphi_0)}$.

Note that  the minimum of $\sqrt{P(\theta,\varphi)} +\sqrt{Q(\theta,\varphi)}$ should occur at the points such that $P(\theta,\varphi)=0$,
$Q(\theta,\varphi)=0$,
\begin{eqnarray}
  \frac{\partial \left(\sqrt{P(\theta,\varphi)} +\sqrt{Q(\theta,\varphi)}\right)}{\partial \theta} &=& 0, \label{differencial-theta}\\
  \frac{\partial \left(\sqrt{P(\theta,\varphi)} +\sqrt{Q(\theta,\varphi)}\right)}{\partial \varphi} &=& 0, \label{differencial-varphi}
\end{eqnarray}
or, the boundary of $\theta$ and $\varphi$. From Eq.(\ref{differencial-theta}) and Eq.(\ref{differencial-varphi}), there are {\small
\begin{eqnarray}
 \frac{\partial P(\theta,\varphi)}{\partial\theta}\frac{\partial Q(\theta,\varphi)}{\partial\varphi}
   -\frac{\partial Q(\theta,\varphi)}{\partial\theta}\frac{\partial P(\theta,\varphi)}{\partial\varphi} &=& 0, \label{diff-theta-varphi}\\
   P(\theta,\varphi)\left(\frac{\partial Q(\theta,\varphi)}{\partial\varphi}\right)^2
   -Q(\theta,\varphi)\left(\frac{\partial P(\theta,\varphi)}{\partial\varphi}\right)^2 &=& 0, \label{diff-varphi}\\
 P(\theta,\varphi)\left(\frac{\partial Q(\theta,\varphi)}{\partial\theta}\right)^2
   -Q(\theta,\varphi)\left(\frac{\partial P(\theta,\varphi)}{\partial\theta}\right)^2 &=& 0. \label{diff-theta}
\end{eqnarray} }

 Let $y=\cot\theta$, $t=\cot\frac{\theta}{2}$, $\theta\in(0,\pi)$, then
$\sin\theta=\frac{1}{\sqrt{y^2+1}}=\frac{2t}{1+t^2}$,$\cos\theta=\frac{y}{\sqrt{1+y^2}}=\frac{t^2-1}{t^2+1}$,$t=y+\sqrt{y^2+1}$,
$t\in(0,+\infty)$. These will be useful throughout the paper.  Next we give controller's optimal measurement
basis (\ref{measurebasis})   for every kind of given quantum channel (\ref{quantumchannel}) satisfying
$a_1a_2a_3a_4\sin\mu =0$. That is, we determine the two parameters $\theta$ and $\varphi$ in measurement basis
(\ref{measurebasis}). To state clearly, we classify the quantum channel (\ref{quantumchannel}) into the
following cases.
 Next we examine quantum channels (\ref{quantumchannel}) with  the following different characterizations, give the maximal
  probability of exact teleportation via any three-qubit state (\ref{quantumchannel}) with $a_1a_2a_3a_4\sin\mu =0$,
  and characterizes the tripartite states that can
 collapse to an EPR pair after Charlie's measurement.

\subsection{$a_1=a_2=a_3=0$, and $a_0a_4\neq 0$}\label{GHZ}

The quantum channel (\ref{quantumchannel}) with three coefficients being 0 is the only one satisfying $a_1=a_2=a_3=0$ and $a_0a_4\neq 0$, since
others are biseparable (one party is not entangled with the other two parties) and can not be used as quantum channel of controlled
teleportation.

It can be seen that
\begin{equation}\label{a123=0min}
\begin{array}{ll}
    & \min\left\{\sqrt{P(\theta,\varphi)} +\sqrt{Q(\theta,\varphi)}\right\} \\
  = & \min\big\{\sqrt{\frac{1}{4}[\cos\theta-(1-2a_0^2)]^2} \\
  & ~~~~~~ +\sqrt{\frac{1}{4}[\cos\theta+(1-2a_0^2)]^2}\big\} \\
  = & |1-2a_0^2|, \\
\end{array}
\end{equation}
 for each $\varphi\in[0,2\pi]$, and all $\theta$ satisfying $|\cos\theta|\leq |1-2a_0^2|$.
It implies that
 \begin{equation}\label{a123=0max}
 p_{\max}=1-|1-2a_0^2|.
\end{equation}

Note that if $\cos\theta=1-2a_0^2\neq 0$ or $\cos\theta=-1+2a_0^2\neq 0$, then $P(\theta,\varphi)=0$ or
$Q(\theta,\varphi)=0$, which means that after Charlie's measurement the state of Alice's particles 2 and 3 can
be an EPR pair with probability $2a_0^2(1-a_0^2)$. Moreover, if $\cos\theta=1-2a_0^2=0$ or
$\cos\theta=-1+2a_0^2= 0$, i.e. $\theta=\frac{\pi}{2}$, $a_0=a_4=\frac{1}{\sqrt{2}}$, then
$P(\theta,\varphi)=Q(\theta,\varphi)=0$, that is, after Charlie measures quantum channel in basis
$\{\frac{|0\rangle+e^{\texttt{i}\varphi}|1\rangle}{\sqrt{2}},
\frac{|0\rangle-e^{\texttt{i}\varphi}|1\rangle}{\sqrt{2}}\}$,  particles 2 and 3 are collapsed to a Bell state
with probability 1.
 It follows that perfect quantum teleportation can be achieved   if  $a_0=\frac{1}{\sqrt{2}}$ (i.e quantum channel is in GHZ state) and $\theta=\frac{\pi}{2}$. That
is, one can send perfect unknown state to another  using GHZ state as a quantum channel via controller's
measurement in the basis $\{\frac{|0\rangle+e^{\texttt{i}\varphi}|1\rangle}{\sqrt{2}},
\frac{|0\rangle-e^{\texttt{i}\varphi}|1\rangle}{\sqrt{2}}\}$, where $\varphi\in[0,2\pi]$.

 \subsection{$a_1=a_4=0$, and $a_0a_2a_3\neq 0$}

Here we consider the quantum channel with coefficients satisfying $a_1=a_4=0$ and $a_0a_2a_3\neq 0$. These
states are called tri-Bell states in \cite{AACJLT}. Note that W-state
\begin{equation}\label{W}
|W\rangle=\frac{1}{\sqrt{3}}(|001\rangle+|010\rangle+|100\rangle)
\end{equation}
is contained here, since the quantum channel (\ref{quantumchannel}) in case of $a_1=a_4=0$ and
$a_0=a_2=a_3=\frac{1}{\sqrt{3}}$ is LOCC equivalent to W-state.

 If $P(\theta,\varphi)=0$, then $\cos\theta=\frac{1 + 2{a_2}{a_3}}{1 - 2a_0^2 + 2{a_2}{a_3}}$, or
 $\cos\theta=\frac{-1 + 2{a_2}{a_3}}{-1 + 2a_0^2 + 2{a_2}{a_3}}$. It is not difficult to prove that
  $|\frac{1 + 2{a_2}{a_3}}{1 - 2a_0^2 + 2{a_2}{a_3}}|>1$, $|\frac{-1 + 2{a_2}{a_3}}{-1 + 2a_0^2 + 2{a_2}{a_3}}|>1$ if $a_2\neq a_3$,
and  $\frac{-1 + 2{a_2}{a_3}}{-1 + 2a_0^2 + 2{a_2}{a_3}}=-1$ if $a_2=a_3$. Thus, $P(\theta,\varphi)=0$ if and
only if $a_2=a_3$, and
  $\theta=\pi$. From (\ref{q-p-relation}), there is  $Q(\theta,\varphi)=0$ if and only if $a_2=a_3$, and
  $\theta=0$. Combining (\ref{q-p-relation}) and (\ref{p-q-relation}), there is
\begin{equation}\label{}
\begin{array}{ll}
 & \left(\sqrt{P(\theta,\varphi)}+\sqrt{Q(\theta,\varphi)}\right)\left|_{P(\theta,\varphi)=0}\right. \\
= & \left(\sqrt{P(\theta,\varphi)}+\sqrt{Q(\theta,\varphi)}\right)\left|_{Q(\theta,\varphi)=0}\right. \\
= & a_0^2.
\end{array}
\end{equation}
Therefore, for the quantum channel (\ref{quantumchannel}) such that $a_1=a_4=0$, $a_2=a_3$  and $a_0a_2a_3\neq
0$, can be collapsed to a Bell state  with probability $p_1=p_2=1-a_0^2$ by Charlie measuring his particle in
the basis $\{|0\rangle, |1\rangle\}$.

  Next we suppose that $P(\theta,\varphi)\neq 0$ and $Q(\theta,\varphi)\neq 0$.
   From (\ref{diff-theta}), we derive that $a_0^4a_2^2a_3^2
  \left( -1 + 2a_0^2 + 4a_2^2a_3^2 \right)\cos\theta\sin^2\theta =0$.  Thus, $\theta=0,\frac{\pi}{2},\pi$. By checking, $\theta=0,\frac{\pi}{2},\pi$ are roots of
  (\ref{differencial-theta}). Note that if $\theta=0, \pi$, then $a_2\neq a_3$ by $P(\theta,\varphi)\neq 0$ and $Q(\theta,\varphi)\neq 0$.    Obviously,
  $\sqrt{P(\theta,\varphi)} +\sqrt{Q(\theta,\varphi)}=a_0^2+|a_2^2-a_3^2|>0$
  if $\theta=0,\pi$ and $a_2\neq a_3$; $\sqrt{P(\theta,\varphi)} +\sqrt{Q(\theta,\varphi)}=\sqrt{1-4a_2^2a_3^2}>0$ if $\theta=\frac{\pi}{2}$.
  Therefore,
\begin{equation}\label{a14=0min}
\begin{array}{ll}
 & \min\left\{\sqrt{P(\theta,\varphi)}
+\sqrt{Q(\theta,\varphi)}\right\} \\
= & \min\big\{a_0^2+|a_2^2-a_3^2|,\sqrt{1-4a_2^2a_3^2}\big\} \\
> & 0.
\end{array}
\end{equation}
That is,  for quantum channel (\ref{quantumchannel}) with $a_1=a_4=0$, $a_0a_2a_3\neq 0$ and
$a_0^2+|a_2^2-a_3^2|<\sqrt{1-4a_2^2a_3^2}$, the controller should choose measurement basis $\{|0\rangle,
|1\rangle\}$; otherwise, he selects measurement basis (\ref{measurebasis}) with $\theta=\frac{\pi}{2}$ (i.e. he
measures in the basis $\{\frac{|0\rangle+e^{\texttt{i}\varphi}|1\rangle}{\sqrt{2}},
\frac{|0\rangle-e^{\texttt{i}\varphi}|1\rangle}{\sqrt{2}}\}$, where $\varphi\in[0,2\pi]$), thus, the controlled
teleportation in Section II can  achieve maximal successful probability $p_{\max}<1$.

 \subsection{one is $a_1=a_2=0$ and $a_0a_3a_4\neq 0$, the other is  $a_1=a_3=0$ and $a_0a_2a_4\neq 0$}

Now we discuss the quantum channel (\ref{quantumchannel}) with the characterization  $a_1=a_2=0$ and $a_0a_3a_4\neq 0$.

 First, $P(\theta,\varphi)\neq 0$ and $Q(\theta,\varphi)\neq 0$. If $P(\theta,\varphi)=0$, then $\cos\theta=\frac{1 - 2a_0^2 - 4a_0^2{a_4}{\sqrt{-a_3^2 }}}{1 - 4a_0^2 +
      4a_0^4 + 4a_0^2a_4^2}$, or  $\cos\theta=\frac{1 - 2a_0^2 + 4a_0^2{a_4}{\sqrt{-a_3^2 }}}{1 - 4a_0^2 +
      4a_0^4 + 4a_0^2a_4^2}$, which are impossible. Thus,  $P(\theta,\varphi)\neq 0$.  By (\ref{q-p-relation}), there is also $Q(\theta,\varphi)\neq
  0$.  The quantum channel (\ref{quantumchannel}) with the characterization  $a_1=a_2=0$ and $a_0a_3a_4\neq 0$
  can never collapse to an EPR pair after Charlie's measurement.  Second, (\ref{diff-theta}) implies $a_0^4(1-2a_0^2)a_4^2a_3^2\cos\theta\sin^2\theta=0$.
  It follows that $\theta=0, \frac{\pi}{2}$,  $\pi$.
  By checking, $\theta=0, \frac{\pi}{2},\pi$ are the roots of
  (\ref{differencial-theta}).  We can get
    \begin{equation}\label{a12=0min}
   \begin{array}{ll}
        \min\{\sqrt{P(\theta,\varphi)} +\sqrt{Q(\theta,\varphi)}\}
         &= \sqrt{P(\frac{\pi}{2},\varphi)} +\sqrt{Q(\frac{\pi}{2},\varphi)}\\
         &  = \sqrt{1-4a_0^2a_4^2}>0.\\
         \end{array}
        \end{equation}
\begin{equation}\label{a12=0max}
     p_{\max}=1-\sqrt{1-4a_0^2a_4^2}<1,
\end{equation}
where  $\theta=\frac{\pi}{2}$.
 Similarly, when $a_1=a_3=0$, and $a_0a_2a_4\neq 0$, we have
\begin{equation}\label{a13=0max}
   p_{\max}=1-\sqrt{1-4a_0^2a_4^2}<1,
\end{equation}
where  $\theta=\frac{\pi}{2}$. That is, the controlled teleportation via quantum channel (\ref{quantumchannel})
satisfying  $a_1=a_2=0$ and $a_0a_3a_4\neq 0$, or $a_1=a_3=0$ and $a_0a_2a_4\neq 0$ can only succeed with
optimal probability $p_{\max}=1-\sqrt{1-4a_0^2a_4^2}<1$.

\subsection{one is $a_2=a_4=0$ and $a_0a_1a_3\neq 0$, the other is $a_3=a_4=0$ and $a_0a_1a_2\neq 0$}

For the quantum channel with characterization either $a_2=a_4=0$ and $a_0a_1a_3\neq 0$, or $a_3=a_4=0$ and $a_0a_1a_2\neq 0$, since
$P(\theta,\varphi)=p_1^2$, $Q(\theta,\varphi)=p_2^2$, so $\min\{\sqrt{P(\theta,\varphi)} +\sqrt{Q(\theta,\varphi)}\}=1$, and $p_{\max}=0$ for
both cases. It can also be seen directly from the quantum channel (\ref{quantumchannel}) with these two characterizations being biseparable
states.

\subsection{$a_2=a_3=0$ and $a_0a_1a_4\neq 0$}\label{nonGHZ}

The quantum channel (\ref{quantumchannel}) with coefficients satisfying  $a_2=a_3=0$, but $a_0a_1a_4\neq 0$, are
extended GHZ states according to the classification in \cite{AACJLT}.

For this kind of quantum channel, we can derive that
\begin{eqnarray}
 P(\theta,\varphi) &=& \frac{1}{4}[ 1- 2a_4^2+ 2a_0a_1\cos(\varphi-\mu)\sin\theta  \nonumber\\
&& -(1-2a_0^2-2a_4^2)\cos\theta]^2,\label{a2=a3=0-P} \\
 Q(\theta,\varphi) &=& \frac{1}{4}[-1+ 2a_4^2+2a_0a_1\cos(\varphi-\mu)\sin\theta \nonumber\\
&& -(1-2a_0^2 -2a_4^2 )\cos\theta]^2, \label{a2=a3=0-Q}
\end{eqnarray}
and
\begin{equation}\label{a23=0max}
\begin{array}{c}
  p_{\max}=1-\min\{\sqrt{P(\theta,\varphi)} +\sqrt{Q(\theta,\varphi)}\}=1-|1-2a_4^2|,
\end{array}
\end{equation}
where $\theta,\varphi$ satisfy $|2a_0a_1\cos(\varphi-\mu)\sin\theta  -( 1 - 2a_0^2 -2a_4^2 )\cos\theta|\leq
|1-2a_4^2|$. Note that the set $S$ of $(\theta,\varphi)$ such that $|2a_0a_1\cos(\varphi-\mu)\sin\theta  -( 1 -
2a_0^2 -2a_4^2 )\cos\theta|\leq |1-2a_4^2|$ is a region, as we can see easily that
$(\frac{\pi}{2},\frac{\pi}{2}+\mu)$, $(\frac{\pi}{2},\frac{3\pi}{2}+\mu)\in S$ in case of $0\leq \mu\leq
\frac{\pi}{2}$, and $(\frac{\pi}{2},\frac{\pi}{2}+\mu)$, $(\frac{\pi}{2},-\frac{\pi}{2}+\mu)\in S$ in case of
$\frac{\pi}{2}\leq \mu\leq \pi$.

From (\ref{a2=a3=0-P}) and (\ref{a2=a3=0-Q}), we have that $ P(\theta,\varphi)= Q(\theta,\varphi)=0$ if and only
if $a_4=\frac{1}{\sqrt{2}}$ and $a_1\cos(\varphi-\mu)\sin\theta+a_0\cos\theta=0$, which means that as long as
Charlie measures the quantum channel
\begin{equation}\label{extended-GHZ-state}
 a_0|000\rangle+a_1e^{\texttt{i}\mu}|100\rangle+a_4|111\rangle, ~~ a_0^2+a_1^2=a_4^2=\frac{1}{2},
\end{equation}
in the basis (\ref{measurebasis}) satisfying $a_1\cos(\varphi-\mu)\sin\theta+a_0\cos\theta=0$, where
$\theta\in[0,\pi]$, $\varphi\in[0,2\pi]$, Alice and Bob can obtain an EPR pair with certainty (i.e. with
probability $p=1$), and Alice teleports her one qubit information to Bob with probability 1 and with unit
fidelity. Note that the state
\begin{equation}\label{extendedGHZ}
 a_0|000\rangle+a_1|100\rangle+a_4|111\rangle, ~~ a_0^2+a_1^2=a_4^2=\frac{1}{2},
\end{equation}
is LOCC equivalent to the state in (\ref{extended-GHZ-state}). That is, the states in (\ref{extendedGHZ}) or
(\ref{extended-GHZ-state}) can be used for perfect controlled quantum teleportation. Clearly, one can achieve
perfect teleportation via these states by the controller making a measurement on his particle using the basis
(\ref{measurebasis}) satisfying $a_1\cos(\varphi-\mu)\sin\theta+a_0\cos\theta=0$. More important, this kind of
states in (\ref{extendedGHZ}) or (\ref{extended-GHZ-state}) are different from GHZ state according to the
classification in \cite{AACJLT}.

\subsection{$a_1=0$ and $a_0a_2a_3a_4\neq 0$}

In this section, we investigate the quantum channel with coefficients having $a_1=0$ and $a_0a_2a_3a_4\neq 0$.

Evidently,
\begin{equation}\label{a1theta=0}
\begin{array}{ll}
& \sqrt{P(0,\varphi)} +\sqrt{Q(0,\varphi)}  \\
= & \sqrt{P(\pi,\varphi)} +\sqrt{Q(\pi,\varphi)} \\
  = & a_0^2+\sqrt{(1-a_0^2)^2-4a_2^2a_3^2} \\
  = &  a_0^2+\sqrt{[a_4^2+(a_2-a_3)^2][a_4^2+(a_2+a_3)^2]},
\end{array}
\end{equation}
and $P(\theta,\varphi)Q(\theta,\varphi)\neq 0$ in case of $\theta=0, \pi$.

We first examine the condition of $P(\theta,\varphi)=0$. If $P(\theta,\varphi)=0$, then
\begin{equation}\label{}
\begin{array}{ll}
\cos\varphi=  & [2\sin\theta( 1 - \cos\theta){a_0}{a_2}{a_3}{a_4}]^{-1}[-\frac{1}{4} + a_2^2a_3^2   \\
 & +a_0^2a_4^2+\cos\theta(\frac{1}{2} - a_0^2 - 2a_2^2a_3^2 ) \\
& -\frac{1}{4}{\cos^2\theta}(1 - 4a_0^2+4a_0^4 - 4a_2^2a_3^2 +4a_0^2a_4^2 )].
\end{array}
\end{equation}
 Note that
\begin{equation}\label{}
\begin{array}{rl}
    z= &[2\sin\theta( 1 - \cos\theta){a_0}{a_2}{a_3}{a_4}]^{-1}[-\frac{1}{4} + a_2^2a_3^2   \\
 & +a_0^2a_4^2+\cos\theta(\frac{1}{2} - a_0^2 - 2a_2^2a_3^2 ) \\
& -\frac{1}{4}{\cos^2\theta}(1 - 4a_0^2+4a_0^4 - 4a_2^2a_3^2 +4a_0^2a_4^2 )] \\
  =  & \frac{-a_0^4t^4 + ( -2a_0^2 +
2a_0^4 + 4a_0^2a_4^2 )t^2 - {( 1 - a_0^2) }^2 + 4a_2^2a_3^2 }{8t{a_0}{a_2}{a_3}{a_4}}
\end{array}
\end{equation}
  tends to $-\infty$ when $t$ tends to 0 and $+\infty$, and $z=-1$ holds only if $a_2=a_3$ and
$t=\frac{a_4}{a_0}$. It follows that $z\leq -1$, where the equality occurs if and only if  $a_2=a_3$ and
$\theta=\theta_0$, where $\cot\frac{\theta_0}{2}=\frac{a_4}{a_0}$ and $0<\theta_0<\pi$. Therefore, only if
$a_2=a_3$, $\varphi=\pi$, $\theta=\theta_0$, there is $P(\theta,\varphi)=0$.

By (\ref{q-p-relation}) and (\ref{p-q-relation}), we have $Q(\theta,\varphi)=0$ iff $a_2=a_3$, $\varphi=0$, and
$\theta=\pi-\theta_0$, and
\begin{equation}\label{}
\begin{array}{ll}
  & \big(\sqrt{P(\theta,\varphi)} +\sqrt{Q(\theta,\varphi)}\big)\big|_{P(\theta,\varphi)=0} \\
 =& \sqrt{Q(\theta_0,\pi)}|_{a_2=a_3} \\
 = & \big(\sqrt{P(\theta,\varphi)} +\sqrt{Q(\theta,\varphi)}\big)\big|_{Q(\theta,\varphi)=0} \\
 = & \sqrt{P(\pi-\theta_0,0)}|_{a_2=a_3} \\
= & \frac{\sqrt{a_0^8 + 2a_0^4a_4^2 - 2a_0^6a_4^2 + a_4^4 - 2a_0^2a_4^4 - 3a_0^4a_4^4 +8a_0^2a_3^2a_4^4 - 4a_3^4a_4^4}}{ a_0^2 + a_4^2} \\
= & \sqrt{1-4a_3^2+4a_3^4-4a_4^2+12a_3^2a_4^2+4a_4^4}.
\end{array}
\end{equation}
 That is, $\sqrt{P(\theta,\varphi)} +\sqrt{Q(\theta,\varphi)}= \sqrt{1-4a_3^2+4a_3^4-4a_4^2+12a_3^2a_4^2+4a_4^4}$
 if $P(\theta,\varphi)=0$ or $Q(\theta,\varphi)=0$.

 It is shown
that  an EPR pair can be obtained with probability $p_1=p_2=\frac{a_0^2(1-a_0^2+a_4^2)}{a_0^2+a_4^2}$  if
$a_2=a_3$ and Charlie using measurement basis (\ref{measurebasis}) with $(\theta,\varphi)=(\theta_0,\pi)$,
 or
$(\theta,\varphi)=(\pi-\theta_0,0)$, where $\cot\frac{\theta_0}{2}=\frac{a_4}{a_0}$ and $0<\theta_0<\pi$.

Next we assume that $P(\theta,\varphi)Q(\theta,\varphi)\neq 0$. From (\ref{diff-theta-varphi})  and (\ref{diff-varphi}),  there are
\begin{equation*}
 \sin\varphi[ 2a_2{a_3}{a_4}\cos\varphi\sin\theta-2a_0^3\cos\theta  + {{{a}}_0} \cos\theta (1 - 2a_4^2) ] =0
\end{equation*}  and
\begin{equation*}
\begin{array}{rc}
\sin \varphi[{a_0}(1- 2a_4^2)\cos\theta{\sin ^2\theta} &  \\
+ 2a_0^3{\cos^3\theta}
 + 2{a_2}{a_3}{a_4}\cos \varphi{\sin^3\theta}] &   =0, \\
\end{array}
\end{equation*}
 respectively.
 By $ 2a_2{a_3}{a_4}\cos\varphi\sin\theta-2a_0^3\cos\theta  + {{{a}}_0} \cos\theta (1 - 2a_4^2) =0$,
 and ${a_0}(1- 2a_4^2)\cos\theta{\sin ^2\theta} + 2a_0^3{\cos^3\theta} + 2{a_2}{a_3}{a_4}\cos \varphi{\sin^3\theta}=0$,
  there are $\theta=\frac{\pi}{2},\varphi=\frac{\pi}{2}, \frac{3\pi}{2}$.  Evidently,
\begin{equation}\label{}
 \begin{array}{ll}
\sqrt{P(\frac{\pi}{2},\frac{\pi}{2})}+\sqrt{Q(\frac{\pi}{2},\frac{\pi}{2})} & =\sqrt{P(\frac{\pi}{2},\frac{3\pi}{2})}
                                                                      +\sqrt{Q(\frac{\pi}{2},\frac{3\pi}{2})}\\
  & =\sqrt{1-4a_2^2a_3^2-4a_0^2a_4^2}. \\
 \end{array}
\end{equation}
  Next we need only to examine the case $
\sin\varphi=0$, i.e. $\varphi=0,\pi,2\pi$, while $P(\theta,\varphi)Q(\theta,\varphi)\neq 0$ and
$\theta\in(0,\pi)$.

Now let us consider the case $\varphi=0$.  From (\ref{diff-theta}), we obtain
\begin{equation}
\begin{array}{ll}
& -4 a_0^3 \left(2 y a_0 a_2 a_3+a_4\right)[4 y^2 a_0^2 a_2 a_3 a_4^2+2 y a_0 (2 a_0^4+ \\
& 2 a_0^2 a_4^2-3 a_0^2-4 a_2^2 a_3^2-a_4^2+1) a_4  \\
& +a_2
   a_3 (2 a_0^2-4 a_0^2 a_4^2+4 a_2^2 a_3^2+2 a_4^2-1)] \\
  = & 0,
\end{array}
\end{equation}
which implies that
\begin{eqnarray}
  y &=& \cot\theta_1=-\frac{a_4}{2a_0a_2a_3}, \\
  y &=& \cot\theta_2=\left\{\begin{array}{lll}
    \frac{a_3(1-2a_3^2-2a_4^2)}{2a_0a_2a_4}, & \texttt{if} & a_2> a_3,\\
   \frac{a_2(1-2a_2^2-2a_4^2)}{2a_0a_3a_4}, & \texttt{if} & a_2<a_3,\\
  \end{array}\right. \\
  y &=& \cot\theta_3=\left\{\begin{array}{lll}
    \frac{a_2(1-2a_2^2-2a_4^2)}{2a_0a_3a_4}, & \texttt{if} & a_2> a_3,\\
   \frac{a_3(1-2a_3^2-2a_4^2)}{2a_0a_2a_4}, & \texttt{if} & a_2<a_3. \\
  \end{array}\right.
\end{eqnarray}
Note that when $a_2=a_3$, then there is
$\cot\theta_2=\cot\theta_3=\cot(\pi-\theta_0)=\frac{1-2a_3^2-2a_4^2}{2a_0a_4}=\frac{a_0^2-a_4^2}{2a_0a_4}$,
which implies that $\theta_2=\theta_3=\pi-\theta_0$ (i.e. $Q(\theta_2,0)=0$) since
$\theta_2,\theta_3,\pi-\theta_0\in (0,\pi)$. Hence $a_2\neq a_3$ in the expression $y=\cot\theta_2$ and
$y=\cot\theta_3$ because of the hypothesis $P(\theta,\varphi)Q(\theta,\varphi)\neq 0$.
 By checking, it can be proved that $\theta_1, \theta_2$  are the roots of
$\frac{\partial\left(\sqrt{P(\theta,0)} +\sqrt{Q(\theta,0)}\right)}{\partial\theta}=0$.  Thus
\begin{widetext}
\begin{equation}\label{a1=0=varphi-min}
\begin{array}{ll}
 & \min\left\{\sqrt{P(\theta,0)} +\sqrt{Q(\theta,0)}\right\} \\
= & \left\{ \begin{array}{lll}
 \min \left\{\sqrt{P(\theta_1,0)} +\sqrt{Q(\theta_1,0)},
 \sqrt{P(\theta_2,0)} +\sqrt{Q(\theta_2,0)},\sqrt{P(0,0)} +\sqrt{Q(0,0)} \right\}, & \texttt{if} & a_2\neq a_3, \\
\min \left\{ \sqrt{P(\pi-\theta_0,0)}\big|_{a_2= a_3},\sqrt{P(\theta_1,0)} +\sqrt{Q(\theta_1,0)},\sqrt{P(0,0)} +\sqrt{Q(0,0)} \right\}, & \texttt{if} & a_2= a_3. \\
\end{array}\right.
\end{array}
\end{equation}
\end{widetext}
For the case $\varphi=\pi$ and the case $\varphi=2\pi$, from (\ref{q-p-relation}), (\ref{p-q-relation}), and
(\ref{0-pi-2pi}), we have
\begin{equation}\label{a1=0-varphi=pi-min}
\begin{array}{ll}
  & \min\{\sqrt{P(\theta,\pi)} +\sqrt{Q(\theta,\pi)}\} \\
 = & \min\{\sqrt{P(\pi-\theta,0)}+\sqrt{Q(\pi-\theta,0)}\} \\
 = & \min\{\sqrt{P(\theta,0)} +\sqrt{Q(\theta,0)}\} \\
  = & \min\{\sqrt{P(\theta,2\pi)} +\sqrt{Q(\theta,2\pi)}\}. \\
\end{array}
\end{equation}

Therefore, if $a_2\neq a_3$, then
\begin{equation}\label{a1=0-min}
\begin{array}{ll}
   &  \min\{\sqrt{P(\theta,\varphi)} +\sqrt{Q(\theta,\varphi)}\} \\
  = & \min\big\{\sqrt{P(0,\varphi)}+\sqrt{Q(0,\varphi)},\sqrt{P(\theta_1,0)} +\sqrt{Q(\theta_1,0)}, \\
   & ~~~~~~~ \sqrt{P(\theta_2,0)} +\sqrt{Q(\theta_2,0)}, \sqrt{P(\frac{\pi}{2},\frac{\pi}{2})} +\sqrt{Q(\frac{\pi}{2},\frac{\pi}{2})} \big\}; \\
\end{array}
\end{equation}
if $a_2=a_3$, then
\begin{equation}\label{a1=0-min}
\begin{array}{ll}
   &  \min\{\sqrt{P(\theta,\varphi)} +\sqrt{Q(\theta,\varphi)}\} \\
  = & \min\big \{\sqrt{P(0,\varphi)}+\sqrt{Q(0,\varphi)},
 \sqrt{P(\pi-\theta_0,0)}\big|_{a_2=a_3}, \\
   &  ~~~~~~~
\sqrt{P(\theta_1,0)} +\sqrt{Q(\theta_1,0)},\sqrt{P(\frac{\pi}{2},\frac{\pi}{2})} +\sqrt{Q(\frac{\pi}{2},\frac{\pi}{2})} \big \}. \\
\end{array}
\end{equation}

\subsection{$a_4=0$ and $a_0a_1a_2a_3\neq 0$}

In this section, we consider the quantum channel (\ref{quantumchannel}) with coefficients  satisfying $a_4=0$ and $a_0a_1a_2a_3\neq 0$.

Note that
\begin{equation*}
 \begin{array}{ll}
& \sqrt{P(0,\varphi)} +\sqrt{Q(0,\varphi)}\\
= & \sqrt{P(\pi,\varphi)} +\sqrt{Q(\pi,\varphi)} \\
= & a_0^2+\sqrt{(1-a_0^2)^2-4a_2^2a_3^2} \\
= &  a_0^2+\sqrt{[a_1^2+(a_2-a_3)^2][a_1^2+(a_2+a_3)^2]},
 \end{array}
\end{equation*}
and $P(\theta,\varphi)Q(\theta,\varphi)\neq 0$ in case of $\theta=0, \pi$.
 Next we suppose that $\sin\theta\neq 0$, that is $\theta\in(0,\pi)$.

We begin with the discussion of the condition such that $P(\theta,\varphi)=0$. We can see that
\begin{equation}\label{a4=0-p}
\begin{array}{ll}
     & P(\theta,\varphi) \\
     = & \frac{1}{4}[ 1 + 2{{{a}}_0}{a_1}\cos(\varphi-\mu)\sin\theta - 2{a_2}{a_3} \\
&  - \cos\theta( 1 - 2a_0^2 - 2{a_2}{a_3})  ]  [ 1 + 2{{{a}}_0}{a_1}\cos(\varphi-\mu)\sin\theta \\
 & + 2{a_2}{a_3} - \cos\theta ( 1 - 2a_0^2 + 2{a_2}{a_3})  ] \\
     = & \frac{1 }{( 1 + t^2)^2}[ 1 + ( t^2-1) a_0^2 + 2t{a_0}{a_1}\cos (\varphi-\mu) -
      2{a_2}{a_3}] \\
&  \times[ 1 + (t^2-1) a_0^2 +
      2t{a_0}{a_1}\cos(\varphi-\mu) + 2{a_2}{a_3} ] \\
    = & 0
    \end{array}
\end{equation}
implies that
\begin{equation}\label{a4=0-cos(varphi-mu)}
\begin{array}{c}
  \cos(\varphi-\mu)=z_1=\frac{-1 + a_0^2 - t^2a_0^2 - 2{a_2}{a_3}}{2t{a_0}{a_1}}, \\
  \cos(\varphi-\mu)=z_2=\frac{-1 + a_0^2 - t^2a_0^2 + 2{a_2}{a_3}}{2t{a_0}{a_1}}.\\
\end{array}
\end{equation}
Note that both $z_1=\frac{-1 + a_0^2 - t^2a_0^2 - 2{a_2}\,{a_3}}{2\,t\,{a_0}\,{a_1}}$  and $z_2=\frac{-1 + a_0^2
- t^2a_0^2 + 2\,{a_2}\,{a_3}}{2\,t\,{a_0}\,{a_1}}$ tend to $-\infty$ when $t\rightarrow 0$  and $t\rightarrow
+\infty$.  Since $z_1\leq -2\sqrt{\frac{1 - a_0^2 + 2\,{a_2}\,{a_3}}{2\,t\,{a_0}\,{a_1}}}\sqrt{\frac{t^2a_0^2
}{2\,t\,{a_0}\,{a_1}}}=-\frac{\sqrt{a_1^2+(a_2+a_3)^2}}{a_1}<-1$, there is no $\varphi$ such that $
\cos(\varphi-\mu)=z_1=\frac{-1 + a_0^2 - t^2a_0^2 - 2\,{a_2}\,{a_3}}{2\,t\,{a_0}\,{a_1}}$. Note that $z_2\leq
-2\sqrt{\frac{1 - a_0^2 -2\,{a_2}\,{a_3}}{2\,t\,{a_0}\,{a_1}}}\sqrt{\frac{t^2a_0^2
}{2\,t\,{a_0}\,{a_1}}}=-\frac{\sqrt{a_1^2+(a_2-a_3)^2}}{a_1}\leq -1$, where  the equality $z_2=-1$ holds iff
$t=\frac{a_1}{a_0}$ and $a_2=a_3$. It follows that  $ \cos(\varphi-\mu)=z_2=\frac{-1 + a_0^2 - t^2a_0^2 +
2\,{a_2}\,{a_3}}{2\,t\,{a_0}\,{a_1}}$ iff $a_2=a_3$, $\varphi=\pi+\mu$, and $\theta=\theta_0$, where
$\cot\frac{\theta_0}{2}=\frac{a_1}{a_0}$. Thus, $P(\theta,\varphi)=0$ if and only if $a_2=a_3$,
$\varphi=\pi+\mu$, and $\theta=\theta_0$.  Therefore,
\begin{equation}\label{}
\begin{array}{ll}
   & \big(\sqrt{P(\theta,\varphi)}
+\sqrt{Q(\theta,\varphi)}\big)\big|_{P(\theta,\varphi)=0} \\
  = & \sqrt{Q(\theta_0,\mu+\pi)}\big|_{a_2=a_3} \\
 = & \frac{\sqrt{\left(a_0^4 + a_1^2 + a_0^2\,a_1^2 - 2\,a_1^2\,a_3^2 \right) \,
       \left( a_0^4 + a_1^2 + a_0^2\,a_1^2 + 2\,a_1^2\,a_3^2 \right) }}{a_0^2 + a_1^2}. \\
\end{array}
\end{equation}

By (\ref{q-p-relation}), we know that $Q(\theta,\varphi)=0$ if and only if $a_2=a_3$, $\varphi=\mu$, and
$\theta=\pi-\theta_0$ and
\begin{equation}\label{}
\begin{array}{rl}
  & \big(\sqrt{P(\theta,\varphi)}
+\sqrt{Q(\theta,\varphi)}\big)\big|_{Q(\theta,\varphi)=0} \\
 = & \sqrt{P(\pi-\theta_0,\mu)}\big|_{a_2=a_3} \\
 = & \big(\sqrt{P(\theta,\varphi)}
+\sqrt{Q(\theta,\varphi)}\big)\big|_{P(\theta,\varphi)=0}. \\
\end{array}
\end{equation}

From above, we can see that the quantum channel with $a_4=0$, $a_2=a_3$, and $a_0a_1a_2a_3\neq 0$ can be
collapsed to an EPR pair with probability $p_1=p_2=\frac{a_0^2(1-a_0^2-a_1^2)}{a_0^2+a_1^2}$  via Charlie's
appropriate measurement (Charlie measures his particle in measurement basis (\ref{measurebasis}) with
$(\theta,\varphi)=(\theta_0,\mu+\pi)$ or $(\theta,\varphi)=(\pi-\theta_0,\mu)$).

In the following we suppose that  $P(\theta,\varphi)Q(\theta,\varphi)\neq 0$.  From (\ref{diff-theta-varphi})
and (\ref{diff-varphi}), we get
\begin{equation}\label{}
\begin{array}{rl}
\sin(\varphi-\mu)[{a_0}\cos\theta + {{{a}}_1}\cos(\varphi-\mu)\sin\theta ] &= 0,\\
 \sin(\varphi-\mu)[{a_0}\cos\theta + {{{a}}_1}\cos(\varphi-\mu)\sin\theta ] &  \\
\times[{\sin^2\theta} + 2a_0^2{\cos^2\theta} +
      2{a_0} {a_1}\cos\theta\cos(\varphi-\mu)\sin\theta] & =0.
\end{array}
\end{equation}
  It follows that
the minimum of $\sqrt{P(\theta,\varphi)} +\sqrt{Q(\theta,\varphi)}$ must occur at either the hyperplane
$\sin(\varphi-\mu)=0$ or the hyperplane ${{{a}}_0}\cos\theta + {{{a}}_1}\cos(\varphi-\mu)\sin\theta =0$ when
$P(\theta,\varphi)Q(\theta,\varphi)\neq 0$ and $\theta\in(0,\pi)$. Now we consider the two hyperplanes.

If ${{{a}}_0}\cos\theta + {{{a}}_1}\cos(\varphi-\mu)\sin\theta =0$,  then $\cos(\varphi-\mu)=-\frac{a_0\cot\theta}{a_1}$,  {\small
\begin{equation}\label{}
\begin{array}{ll}
   & \big(\sqrt{P(\theta,\varphi)}
+\sqrt{Q(\theta,\varphi)}\big)\big|_{\cos(\varphi-\mu)=-\frac{a_0\cot\theta}{a_1}} \\
  = & \sqrt{1-4a_2^2a_3^2}. \\
\end{array}
\end{equation}  }

For $\sin(\varphi-\mu)=0$, i.e. $\varphi=\mu,\mu+\pi$ in case of $\mu\in(0,\pi]$, or $\varphi=0,\pi,2\pi$ in
case of $\mu=0$, we first investigate the case $\varphi=\mu$. From (\ref{diff-theta}), we have
\begin{equation}\label{}
\begin{array}{ll}
  ({a_0}\cos\theta + {a_1}\sin\theta)[2a_0{a_1}({a}_0^2- a_1^2)\sin 2\theta
  - 2a_0^2({\sin^2\theta} & \\
   +  2a_1^2\cos 2\theta ) +
      {\sin^2\theta}(1 - 2a_1^2 - 4a_2^2a_3^2 )]=0, &
\end{array}
\end{equation}
 that is, ${a_0}\cos\theta + {a_1}\sin\theta=0$, or
 $2a_0{a_1}({a}_0^2- a_1^2)\sin 2\theta - 2a_0^2({\sin^2\theta} +  2{{a}}_1^2\cos 2\theta ) +
      {\sin^2\theta}(1 - 2a_1^2 - 4a_2^2a_3^2 )=0$.

 From ${a_0}\cos\theta + {a_1}\sin\theta=0$,
 there is
\begin{equation*}\label{}
\cot\theta_1=-\frac{a_1}{a_0}.
\end{equation*}
 Clearly, $\theta_1$ is a root of (\ref{differencial-theta}). From
\begin{equation}\label{}
\begin{array}{rl}
  2a_0{a_1}({a}_0^2- a_1^2)\sin 2\theta - 2a_0^2({\sin^2\theta} +  2{{a}}_1^2\cos 2\theta ) & \\
   + {\sin^2\theta}(1 - 2a_1^2 - 4a_2^2a_3^2 ) & =0, \\
\end{array}
\end{equation}
 there is
 \begin{eqnarray*}
 \cot\theta_2 &=& \frac{a_0^2-a_1^2-|a_2^2-a_3^2|}{2a_0a_1}, \\
\cot\theta_3 &=& \frac{a_0^2-a_1^2+|a_2^2-a_3^2|}{2a_0a_1},
 \end{eqnarray*}
where $a_2\neq a_3$.  (Since when $a_2=a_3$ there is $\theta_2=\theta_3=\pi-\theta_0$, i.e.
$Q(\pi-\theta_0,\mu)=0$, which contradict with the hypothesis $P(\theta,\varphi)Q(\theta,\varphi)\neq 0$).  By
checking, we see $\theta_3$ is a root of $\frac{\partial\big(\sqrt{P(\theta,\mu)}
+\sqrt{Q(\theta,\mu)}\big)}{\partial\theta}=0$, while $\theta_2$ is not. Hence,
\begin{widetext}
\begin{equation}
\begin{array}{ll}
 &  \min\{\sqrt{P(\theta,\mu)}+\sqrt{Q(\theta,\mu)}\} \\
= & \left\{\begin{array}{lll}
 \min\left\{ \sqrt{P(\theta_1,\mu)}+\sqrt{Q(\theta_1,\mu)}, \sqrt{P(\theta_3,\mu)}+\sqrt{Q(\theta_3,\mu)},
 \sqrt{P(0,\mu)}+\sqrt{Q(0,\mu)} \right\}, & \texttt{if} & a_2\neq a_3,\\
  \min\left\{ \sqrt{P(\theta_1,\mu)}+\sqrt{Q(\theta_1,\mu)}, \sqrt{P(\pi-\theta_0,\mu)}|_{a_2= a_3},
  \sqrt{P(0,\mu)}+\sqrt{Q(0,\mu)} \right\},  & \texttt{if} & a_2= a_3. \\
\end{array} \right.
\end{array}
\end{equation}
\end{widetext}
For $\varphi=\mu+\pi,\mu+2\pi$,  there is
\begin{equation}\label{mu-pi-2pi}
\begin{array}{ll}
  & \min\{\sqrt{P(\theta,\mu+\pi)}+\sqrt{Q(\theta,\mu+\pi)}\} \\
= & \min\{\sqrt{P(\pi-\theta,\mu)}+\sqrt{Q(\pi-\theta,\mu)}\} \\
= & \min\{\sqrt{P(\theta,\mu)}+\sqrt{Q(\theta,\mu)}\} \\
= & \min\{\sqrt{P(\theta,\mu+2\pi)}+\sqrt{Q(\theta,\mu+2\pi)}\}.
\end{array}
\end{equation}
from (\ref{p-q-relation}) and (\ref{q-p-relation}).

Thus, if $a_2=a_3$, then
\begin{equation}\label{a4=0-min}
\begin{array}{ll}
 & \min\big\{\sqrt{P(\theta,\varphi)}+\sqrt{Q(\theta,\varphi)}\big\} \\
  = & \min\big\{a_0^2+\sqrt{(1-a_0^2)^2-4a_2^2a_3^2}, \sqrt{1-4a_2^2a_3^2},\\
   & ~~~~~~~ \sqrt{P(\theta_1,\mu)}+\sqrt{Q(\theta_1,\mu)}, \sqrt{Q(\theta_0,\pi+\mu)}\big|_{a_2=a_3}\big\}; \\
\end{array}
\end{equation}
 if $a_2\neq a_3$, then
\begin{equation}\label{a4=0-min}
\begin{array}{ll}
 & \min\big\{\sqrt{P(\theta,\varphi)}+\sqrt{Q(\theta,\varphi)}\big\} \\
  = & \min\big\{a_0^2+\sqrt{(1-a_0^2)^2-4a_2^2a_3^2},  \sqrt{1-4a_2^2a_3^2},\\
   & ~~~~~~~  \sqrt{P(\theta_1,\mu)}+\sqrt{Q(\theta_1,\mu)},
        \sqrt{P(\theta_3,\mu)}+\sqrt{Q(\theta_3,\mu)} \big\}. \\
\end{array}
\end{equation}
Here $\cot\frac{\theta_0}{2}=\frac{a_1}{a_0}$, $\cot\theta_1=- \frac{a_1}{a_0}$, $ \cot\theta_3=
\frac{a_0^2-a_1^2+|a_2^2-a_3^2|}{2a_0a_1}$.

\subsection{One is $a_2=0$ and $a_0a_1a_3a_4\neq 0$, the other is $a_3=0$ and $a_0a_1a_2a_4\neq 0$}

We now discuss the quantum channel (\ref{quantumchannel}), the coefficients of which satisfying  $a_2=0$, but $a_0a_1a_3a_4\neq 0$.

When $\sin\theta=0$, i.e. $\theta=0, \pi$, then
\begin{equation}\label{}
\begin{array}{ll}
&  \sqrt{P(0,\varphi)}+\sqrt{Q(0,\varphi)} \\
= & \sqrt{P(\pi,\varphi)}+\sqrt{Q(\pi,\varphi)} \\
=  & a_0^2+\sqrt{(1-a_0^2)^2-4a_1^2a_4^2} \\
= & a_0^2+\sqrt{[a_3^2+(a_1-a_4)^2][a_3^2+(a_1+a_4)^2]},\\
\end{array}
\end{equation}
and $P(\theta,\varphi)Q(\theta,\varphi)\neq 0$ in case of $\theta=0, \pi$. Next we suppose that $\sin\theta\neq 0$.

First, we prove that $P(\theta,\varphi)Q(\theta,\varphi) \neq 0$. If  $P(\theta,\varphi)=0$, then
\begin{eqnarray*}
&& \cos(\varphi-\mu) \\
&=& \frac{1}{2a_0^2a_1^2\sin^2\theta}\big[-a_0^3a_1\sin 2\theta +8\sqrt{-a_0^2a_1^2a_3^2a_4^2\cos^2\frac{\theta}{2}\sin^6\frac{\theta}{2}} \\
&& -4a_0a_1(1-2a_4^2)\cos\frac{\theta}{2}\sin^3\frac{\theta}{2}\big], \\
&& \cos(\varphi-\mu)  \\
&=& -\frac{1}{2a_0^2a_1^2\sin^2\theta}\big[a_0^3a_1\sin 2\theta +8\sqrt{-a_0^2a_1^2a_3^2a_4^2\cos^2\frac{\theta}{2}\sin^6\frac{\theta}{2}} \\
&& +4a_0a_1(1-2a_4^2)\cos\frac{\theta}{2}\sin^3\frac{\theta}{2}\big].
\end{eqnarray*}
But $8\sqrt{-a_0^2a_1^2a_3^2a_4^2\cos^2(\frac{\theta}{2})\sin^6(\frac{\theta}{2})}$   is an imaginary number, so
$P(\theta,\varphi)\neq 0$. From (\ref{q-p-relation}), there is also $Q(\theta,\varphi) \neq 0$. That is, no
matter what kind of measurement basis Charlie choose, Alice and Bob can never share an EPR pair  after his
measurement.

Second, the minimum of $\sqrt{P(\theta,\varphi)}+\sqrt{Q(\theta,\varphi)}$ should occur at the point satisfying (\ref{differencial-theta}) and
(\ref{differencial-varphi}), or $\theta=0$. From (\ref{diff-theta-varphi}) and (\ref{diff-varphi}), there are
\begin{equation}
\sin(\varphi-\mu)[a_0\cos\theta +a_1\cos(\varphi-\mu)\sin\theta]=0,
\end{equation}
and
\begin{equation}\label{}
\begin{array}{ll}
 & \sin(\varphi-\mu)[a_0\cos\theta+a_1\cos(\varphi-\mu)\sin\theta][2a_0^2\cos^2\theta \\
 & +a_0a_1\cos(\varphi-\mu)\sin2\theta+(1-2a_4^2)\sin^2\theta]=0.
\end{array}\end{equation}
It follows that the minimum of $\sqrt{P(\theta,\varphi)}+\sqrt{Q(\theta,\varphi)}$ should occur at the point
such that $\sin(\varphi-\mu)=0$, or $a_0\cos\theta +a_1\cos(\varphi-\mu)\sin\theta=0$, or $\theta=0$.

From $a_0\cos\theta +a_1\cos(\varphi-\mu)\sin\theta=0$, we know that
$\cos(\varphi-\mu)=-\frac{a_0\cot\theta}{a_1}$, and
\begin{equation}\label{}
 \begin{array}{cl}
    & \left(\sqrt{P(\theta,\varphi)}+\sqrt{Q(\theta,\varphi)}\right)\big|_{\{\theta\in(0,\pi),\cos(\varphi-\mu)=-\frac{a_0\cot\theta}{a_1}\}} \\
   =  & \sqrt{1-4a_0^2a_4^2-4a_1^2a_4^2}.
 \end{array}
\end{equation}

 For the case $\sin(\varphi-\mu)=0$, since equality (\ref{mu-pi-2pi}) holds,  we need only to
consider the case $\varphi=\mu$. From (\ref{diff-theta}),we derive
\begin{equation}\label{}
\begin{array}{cl}
  & \sin3\theta \left(2\, a_0^2\, {a_1}-a_1 + 4\, a_0^4\, {a_1} + 2\, a_1^3 - 12\, a_0^2\, a_1^3 \right) \\
& + \sin\theta \left(3\,
{a_1} - 6\, a_0^2\, {a_1} + 4\, a_0^4\, {a_1} - 6\, a_1^3 + 4\, a_0^2\, a_1^3 \right) \\
 & + \cos\theta \left({a_0} - 2\, a_0^3 - 2\, {a_0}\, a_1^2 -
        4\, a_0^3\, a_1^2 - 4\, {a_0}\, a_1^4 \right) \\
& + \cos3 \theta\left( 2\, a_0^3-{a_0} + 2\, {a_0}\, a_1^2 -
      12\, a_0^3\, a_1^2 + 4\, {a_0}\, a_1^4 \right) \\
  =& \{-16\cot^3\theta\,a_0^3\,a_1^2 + 16\cot^2\theta\,a_0^2\,{a_1}\,\left( a_0^2 - 2\,a_1^2 \right) \\
&   +  4\cot\theta {a_0}\,\left[ 1 - 2\,a_1^2 - 4\,a_1^4 + 2a_0^2( 4\,a_1^2-1)  \right]\\
& +  4\,\left(1 - 2\,a_0^2 \right){a_1}\left( 1 - 2\,a_1^2 \right)\}\sin^3\theta\\
= & 0,
\end{array}
\end{equation}
which follows that
\begin{eqnarray*}
 \cot\theta=\cot\theta_1  &=& -\frac{a_1}{a_0}, \\
  \cot\theta=\cot\theta_2 &=& \frac{1-2a_1^2}{2a_0a_1}, \\
 \cot\theta=\cot\theta_3 &=& \frac{-1+2a_0^2}{2a_0a_1},
\end{eqnarray*}
where $\theta_1$ and $\theta_2$ are the roots of equation
$\frac{\partial(\sqrt{P(\theta,\mu)}+\sqrt{Q(\theta,\mu)})}{\partial\theta}=0$, while $\theta_3$ is not.  It is
not difficult to obtain
 \begin{equation}\label{}
 \begin{array}{cl}
& \min\big\{\sqrt{P(\theta,\mu)}+\sqrt{Q(\theta,\mu)}\big\} \\
 = & \min\big\{\sqrt{P(\theta_1,\mu)}+\sqrt{Q(\theta_1,\mu)}, \sqrt{P(\theta_2,\mu)}+\sqrt{Q(\theta_2,\mu)}, \\
& ~~~~~~~~ \sqrt{P(0,\mu)}+\sqrt{Q(0,\mu)} \big\}.
  \end{array}
\end{equation}

Therefore,  for quantum channel (\ref{quantumchannel}) with  $a_2=0$ and $a_0a_1a_3a_4\neq 0$,  there is
\begin{equation}\label{a2=0-min}
\begin{array}{cl}
  & \min\{\sqrt{P(\theta,\varphi)}+\sqrt{Q(\theta,\varphi)}\} \\
 = & \min\Big\{\sqrt{P(0,\varphi)}+\sqrt{Q(0,\varphi)},\sqrt{1-4a_0^2a_4^2-4a_1^2a_4^2},\\
& ~~~~~~~ \sqrt{P(\theta_1,\mu)}+\sqrt{Q(\theta_1,\mu)}, \sqrt{P(\theta_2,\mu)}+\sqrt{Q(\theta_2,\mu)} \Big\}.
\end{array}
\end{equation}
Similarly, for quantum channel (\ref{quantumchannel}) with $a_3=0$ and $a_0a_1a_2a_4\neq 0$  we also obtain
\begin{equation}\label{a3=0-min}
\begin{array}{cl}
  & \min\{\sqrt{P(\theta,\varphi)}+\sqrt{Q(\theta,\varphi)}\} \\
  = &  \min\Big\{a_0^2+\sqrt{(1-a_0^2)^2-4a_1^2a_4^2},  \sqrt{1-4a_0^2a_4^2-4a_1^2a_4^2},\\
   & ~~~~~~ \sqrt{P(\theta_1,\mu)}+\sqrt{Q(\theta_1,\mu)}, \sqrt{P(\theta_2,\mu)}+\sqrt{Q(\theta_2,\mu)} \Big\}.
\end{array}
\end{equation}
Here $\cot\theta_1=-\frac{a_1}{a_0}$.

\subsection{$\mu=0$ and $a_0a_1a_2a_3a_4\neq 0$}

In this section, we investigate the quantum channel (\ref{quantumchannel}) with $\mu=0$ and $a_0a_1a_2a_3a_4\neq
0$. Evidently,
\begin{equation}\label{}
\begin{array}{ll}
 & \sqrt{P(0,\varphi)}+\sqrt{Q(0,\varphi)} \\
 & =\sqrt{P(\pi,\varphi)}+\sqrt{Q(\pi,\varphi)} \\
  & =a_0^2+\sqrt{(1-a_0^2)^2-4(a_2a_3-a_1a_4)^2} \\
 & = a_0^2+\sqrt{[(a_2-a_3)^2+(a_1+a_4)^2][(a_2+a_3)^2+(a_1-a_4)^2]} \\
\end{array}
\end{equation}
 and $P(\theta,\varphi)Q(\theta,\varphi)\neq 0$ in case of $\theta=0, \pi$.
 In the following, we suppose that $\theta\neq 0, \pi$, i.e.
$\theta\in(0,\pi)$.

We first consider the case $P(\theta,\varphi)=0$. If $P(\theta,\varphi)=0$, then
\begin{eqnarray*}
 \cos\varphi &=&  z_1\\
   &=& \frac{1}{2\,t\,{a_0}\,a_1^2}\{- t^2\,a_0^2\,{a_1}-( 1 - a_0^2){a_1} \\
& &  -2(a_2a_3-a_1a_4)a_4-2[t^2\,a_0^2\,{a_1}\,{a_2}\,{a_3}\,{a_4} +  \\
&&  ( {a_2}\,{a_3} - {a_1}\,{a_4})(a_1 a_3 + a_2 a_4)(a_1 a_2 + a_3 a_4)]^{\frac{1}{2}}\}, \\
  \cos\varphi &=&  z_2 \\
  &=& \frac{1}{2\,t\,{a_0}\,a_1^2}\{- t^2\,a_0^2\,{a_1}  -( 1 - a_0^2){a_1} \\
& &  -2(a_2a_3-a_1a_4)a_4+2[t^2\,a_0^2\,{a_1}\,{a_2}\,{a_3}\,{a_4} +   \\
&& ( {a_2}\,{a_3} - {a_1}\,{a_4})(a_1 a_3 + a_2 a_4)(a_1 a_2 + a_3 a_4)]^{\frac{1}{2}}\}. \\
 \end{eqnarray*}
Note that  if
              ${a_1}\,{a_4}-{a_2}\,{a_3}>0$, then $t\in [t_0,+\infty)$; if
 ${a_1}\,{a_4}-{a_2}\,{a_3}\leq 0$, then $t\in (0,+\infty)$. Here,
 $t_0=\sqrt{\frac{\left({a_1}\,{a_4} - {a_2}\,{a_3}\right)
            (a_1 a_3 + a_2 a_4)(a_1 a_2 + a_3 a_4) }{a_0^2\,{a_1}\,{a_2}\,{a_3}\,{a_4}}}$.
 Obviously, both  $z_1$  and $z_2$ go to $-\infty$ when $t\rightarrow +\infty$.

We can prove that $z=z_1= \frac{1}{2\,t\,{a_0}\,a_1^2}\{- t^2\,a_0^2\,{a_1}-( 1 -
a_0^2){a_1}-2(a_2a_3-a_1a_4)a_4 -2[t^2\,a_0^2\,{a_1}\,{a_2}\,{a_3}\,{a_4} +
           \left( {a_2}\,{a_3} - {a_1}\,{a_4} \right) (a_1 a_3 + a_2 a_4)(a_1 a_2 + a_3 a_4)]^{\frac{1}{2}}\} $ has no intersection point with the straight line
 $z=-1$. Since $z_1$ is a continuous function of $t$ and tends to $-\infty$ when $t$ tends to $+\infty$, so $z_1<-1$. Thus, there is no
 $\varphi$ satisfying $\cos\varphi=z_1$. Let us look at $z=z_2$. $z_2=-1$ implies that {\small
\begin{eqnarray}
   && [- t^2\,a_0^2\,{a_1}-( 1 - a_0^2){a_1} -2(a_2a_3-a_1a_4)a_4 +2ta_0a_1^2]^2 \nonumber\\
 && -4\{t^2\,a_0^2\,{a_1}\,{a_2}\,{a_3}\,{a_4} +
           \left( {a_2}\,{a_3} - {a_1}\,{a_4} \right) \nonumber\\
&& \times (a_1 a_3 + a_2 a_4)(a_1 a_2 + a_3 a_4)\} \nonumber\\
   &=& a_1^2\,\left[ - t^2\,a_0^2 + 2\,t\,{a_0}({a_1} + {a_4}) -1 + a_0^2 + 2\,{a_2}\,{a_3}-
    2\,{a_1}\,{a_4} \right] \nonumber\\
&& \times  \left[ - t^2\,a_0^2 + 2\,t\,{a_0}({a_1} -
    {a_4}) -1 + a_0^2  - 2\,{a_2}\,{a_3}+ 2\,{a_1}\,{a_4} \right] \nonumber\\
&=& 0,
\end{eqnarray} }
 which follows that
 \begin{eqnarray*}
&&  t=t_1=\frac{a_1-a_4-\sqrt{-(a_2+a_3)^2}}{a_0},\\
 & & t=t_2=\frac{a_1-a_4+\sqrt{-(a_2+a_3)^2}}{a_0}, \\
&&  t=t_3=\frac{a_1+a_4-\sqrt{-(a_2-a_3)^2}}{a_0},\\
 & & t=t_4=\frac{a_1+a_4+\sqrt{-(a_2-a_3)^2}}{a_0}.
 \end{eqnarray*}
Here $t_1$ and $t_2$ are imaginary numbers, while $t_3$ and $t_4$ are real numbers only if $a_2=a_3$. By
checking, when  $a_2=a_3$, $t_3=t_4=\frac{a_1+a_4}{a_0}$ is a root of equation $z_2=-1$, and the maximum point
of function $z=z_2$. Since $z=z_2$ is a continuous function of $t$, tends to $-\infty$ when $t$ tends to
$+\infty$, and has only one intersection point with straight line $z=-1$, there must be $z_2\leq -1$, where the
equality holds iff $a_2=a_3$  and $t=t_3=t_4=\frac{a_1+a_4}{a_0}$. It means that $P(\theta,\varphi)=0$  iff
$a_2=a_3$, $\varphi=\pi$, and $\theta=\theta_0$, where $\cot\frac{\theta_0}{2}=\frac{a_1+a_4}{a_0}$. From
(\ref{q-p-relation}), we can show that $Q(\theta,\varphi)=0$ iff $a_2=a_3$, $\varphi=0$, and
$\theta=\pi-\theta_0$. Therefore
\begin{eqnarray*}
   && (\sqrt{P(\theta,\varphi)}+\sqrt{Q(\theta,\varphi)})\big|_{P(\theta,\varphi)=0} \\
   &=& \sqrt{Q(\theta_0,\pi)}\big|_{a_2=a_3} \\
   &=& \sqrt{P(\pi-\theta_0,0)}\big|_{a_2=a_3} \\
   &=& (\sqrt{P(\theta,\varphi)}+\sqrt{Q(\theta,\varphi)})\big|_{Q(\theta,\varphi)=0}\\
   &=& \sqrt{(1-2a_4^2)^2+4a_3^2(-1+a_1^2+a_3^2+2a_1a_4+3a_4^2)}.
\end{eqnarray*}

It is shown that  the quantum channel (\ref{quantumchannel}) with $\mu=0$, $a_0a_1a_2a_3a_4\neq 0$, and
$a_2=a_3$, collapses to an EPR pair with probability
$p_1=p_2=\frac{a_0^2(1-a_0^2+3a_1^2+4a_1a_4+a_4^2)}{a_0^2+(a_1+a_4)^2}$  after Charlie's measurement in the
basis (\ref{measurebasis}) with either $\varphi=\pi$ and $\theta=\theta_0$, or $\varphi=0$ and
$\theta=\pi-\theta_0$ in case of $a_2=a_3$.

In the following we suppose that $P(\theta,\varphi)Q(\theta,\varphi)\neq 0$. By (\ref{diff-theta-varphi}), there is
\begin{equation}\label{mu=0-diff-theta-varphi}
\begin{array}{cl}
 & \sin \varphi \{2\cos\varphi(a_2a_3 - a_1a_4) (a_1 a_3 + a_2 a_4)(a_1 a_2 + a_3 a_4) \\
& -ya_0[a_2a_3a_4(1+2a_1^2-2a_2^2-2a_3^2) \\
& -2a_1(a_2^2a_3^2-a_2^2a_4^2-a_3^2a_4^2)]\} \\
 = &  0.
\end{array}
\end{equation}
Note that $a_2a_3a_4(1+2a_1^2-2a_2^2-2a_3^2)-2a_1(a_2^2a_3^2-a_2^2a_4^2-a_3^2a_4^2)=a_1a_4^2\neq 0$ if
$a_2a_3=a_1a_4$. Therefore,
$$a_2a_3a_4(1+2a_1^2-2a_2^2-2a_3^2)-2a_1(a_2^2a_3^2-a_2^2a_4^2-a_3^2a_4^2)=0$$ implies that $\cos
  \varphi=0$. Thus, (\ref{mu=0-diff-theta-varphi}) implies that $\sin \varphi=0$, $\cos \varphi=0$, or
{\small   \begin{equation}\label{mu=0-y}
   y=\frac{2\,\cos \varphi( {a_2}{a_3} - {a_1}{a_4}) [ (a_1^2+a_4^2){a_2}{a_3} +a_1(a_2^2+a_3^2)a_4
]}{a_0[a_2a_3a_4(1+2a_1^2-2a_2^2-2a_3^2)-2a_1(a_2^2a_3^2-a_2^2a_4^2-a_3^2a_4^2)]},
\end{equation} }
in case of $\sin 2\varphi\neq 0$.

From (\ref{diff-varphi}), there is
\begin{equation}\label{mu=0-diff-varphi}
 \begin{array}{l}
 4  y^3 a_0^3 (a_2 a_3-a_1 a_4) (a_1 a_3 + a_2 a_4)(a_1 a_2 + a_3 a_4)\\
  +8 y^2 a_0^2 a_1 \cos\varphi  (a_2 a_3-a_1 a_4)(a_1 a_3 + a_2 a_4)(a_1 a_2 + a_3 a_4) \\
 + y{a_0}\{ 4a_1^2{\cos^2\varphi}(a_2 a_3-a_1 a_4) (a_1 a_3 + a_2 a_4)(a_1 a_2 + a_3 a_4)\\
   +  ( {a_1} + 2{a_2}{a_3}{a_4} - 2{a_1}a_4^2 )
      [a_2 a_3 (-2 a_4^3-4 a_1^2 a_4 \\
   +a_4 +2 a_1 a_2 a_3)-2 a_1 a_4^2 (a_2^2+a_3^2) ] \}\\
   +2\cos\varphi({a_1} + 2{a_2}{a_3}{a_4} - 2{a_1}a_4^2 ) [a_2^2 a_3^2 (a_1^2+a_4^2) \\
 +a_1 a_2 a_3 a_4 (1-2 a_1^2-2 a_4^2)-a_1^2 a_4^2 (a_2^2+a_3^2)] \\
 =0.
\end{array}
\end{equation}
Combining (\ref{mu=0-y}) and  (\ref{mu=0-diff-varphi}), we obtain that
\begin{equation}\label{}
\begin{array}{ll}
& \cos\varphi(a_1+2a_2a_3a_4-2a_1a_4^2)\{a_0^2[2a_1(a_2^2a_3^2-a_2^2a_4^2\\
& -a_3^2a_4^2)-a_2a_3a_4(1+2a_1^2-2a_2^2-2a_3^2)]^2   \\
 & +4\cos^2\varphi( {a_2}{a_3} - {a_1}{a_4})^2 (a_1a_3+a_2a_4)^2(a_1a_2+a_3a_4)^2 \} \\
 & =0,
\end{array}
\end{equation}
which implies that $\cos \varphi=0$.

Next we only need to examine the two cases $\sin \varphi=0$, and $\cos \varphi=0$, that is,
$\varphi=0,\frac{\pi}{2}, \pi, \frac{3\pi}{2},2\pi$.

Now let look at the hyperplane $\varphi=0$. From (\ref{diff-theta}), we have
\begin{equation}\label{mu=0-diff-theta}
\begin{array}{ll}
  & [2y{a_0}( {a_2}{a_3} - {a_1}{a_4} ) +2{a_1}{a_2}{a_3} + {a_4} - 2a_1^2{a_4}]\\
& \times\{ {a_2}{a_3} - 2a_0^2{a_2}{a_3} - 2a_1^2{a_2}{a_3} + 4a_0^2a_1^2{a_2}{a_3} \\
   &  - 4a_2^3a_3^3 - 2{a_1}{a_4} + 6a_0^2{a_1}{a_4} -
    4a_0^4{a_1}{a_4} + 2a_1^3{a_4}  \\
&  - 4a_0^2a_1^3{a_4} +  8{a_1}a_2^2a_3^2{a_4}- 2{a_2}{a_3}a_4^2 + 4a_0^2{a_2}{a_3}a_4^2  \\
   &  - 4a_1^2{a_2}{a_3}a_4^2 + 2{a_1}a_4^3 - 4a_0^2{a_1}a_4^3\\
& - 4y^2a_0^2(a_1 a_3 + a_2 a_4)(a_1 a_2 + a_3 a_4) \\
&  - 2y{a_0}( -2a_0^2{a_1}{a_2}{a_3} + 2a_1^3{a_2}{a_3} + {a_4} - 3a_0^2{a_4} \\
& +  2a_0^4{a_4} + a_1^2{a_4} - 2a_1^4{a_4} - 4a_2^2a_3^2{a_4} +
       6{a_1}{a_2}{a_3}a_4^2   \\
& - a_4^3 + 2a_0^2a_4^3 - 2a_1^2a_4^3 )\}\\
=& 0.
\end{array}
\end{equation}
First, we consider the case $a_1a_4-a_2a_3\neq 0$. In this case, (\ref{mu=0-diff-theta}) has the following
solution
 \begin{eqnarray*}
  y=\cot\theta_1&=& \frac{2a_1a_2a_3+a_4-2a_1^2a_4}{2a_0(a_1a_4-a_2a_3)},\\
          && \texttt{where if}~ a_2=a_3, \texttt{then}~ a_1a_4-a_2a_3\neq \frac{1}{2};\\
  y=\cot\theta_2&=& \frac{-2a_1a_2a_4+a_3(1-2a_1^2-2a_3^2-2a_4^2)}{2a_0(a_1a_3+a_2a_4)},\\
          && \texttt{where}~ a_2\neq a_3;\\
 y=\cot\theta_3 &=& \frac{-2a_1a_3a_4+a_2(1-2a_1^2-2a_2^2-2a_4^2)}{2a_0(a_1a_2+a_3a_4)}, \\
          && \texttt{where}~  a_2\neq a_3.
\end{eqnarray*}
By checking, we show that $\theta_1$ is a root of
$\frac{\partial(\sqrt{P(\theta,0)}+\sqrt{Q(\theta,0)})}{\partial\theta}\Big|_{a_1a_4-a_2a_3\neq 0}=0$, while
$\theta_2$ is a root only if $a_2>a_3$, and $\theta_3$ is a root only if $a_2<a_3$.

Second, we consider the case  $a_1a_4-a_2a_3= 0$. From (\ref{mu=0-diff-theta}), we get
\begin{eqnarray*}
  y_4=\cot\theta_4 = & \frac{(1-2a_0^2)(-1+a_0^2-a_1^2+a_4^2+|a_2^2-a_3^2|)}{4a_0(1-a_0^2)a_1}, & \\
                   & \texttt{where} ~  a_2\neq a_3; & \\
  y_5=\cot\theta_5 = & \frac{(1-2a_0^2)(-1+a_0^2-a_1^2+a_4^2-|a_2^2-a_3^2|)}{4a_0(1-a_0^2)a_1},& \\
                     &  \texttt{where} ~ a_2\neq a_3. &
\end{eqnarray*}
By checking, we know that only $\theta_4$ is a root of
$\frac{\partial(\sqrt{P(\theta,0)}+\sqrt{Q(\theta,0)})}{\partial\theta}\left|_{a_1a_4-a_2a_3= 0}\right.=0$.

 Therefore,
 \begin{widetext}
\begin{equation}\label{}
\begin{array}{ll}
 & \min\{\sqrt{P(\theta,0)}+\sqrt{Q(\theta,0)}\} \\
= & \left \{\begin{array}{lllll}
\min\left\{\sqrt{P(\theta_1,0)}+\sqrt{Q(\theta_1,0)}, \sqrt{P(\theta_2,0)}+\sqrt{Q(\theta_2,0)}, \sqrt{P(0,0)}+\sqrt{Q(0,0)} \right\}, & \texttt{if} & a_2> a_3 & \texttt{and} & a_1a_4-a_2a_3\neq 0, \\
\min\left\{\sqrt{P(\theta_1,0)}+\sqrt{Q(\theta_1,0)}, \sqrt{P(\theta_3,0)}+\sqrt{Q(\theta_3,0)}, \sqrt{P(0,0)}+\sqrt{Q(0,0)} \right\}, & \texttt{if} & a_2<a_3 & \texttt{and} & a_1a_4-a_2a_3\neq 0, \\
\min\left\{ \sqrt{P(\theta_4,0)}+\sqrt{Q(\theta_4,0)}, \sqrt{P(0,0)}+\sqrt{Q(0,0)} \right\},  & \texttt{if} & a_2\neq a_3 & \texttt{and} & a_1a_4-a_2a_3= 0,\\
\min\left\{ \sqrt{P(\pi-\theta_0,0)}|_{a_2= a_3}, \sqrt{P(\theta_1,0)}+\sqrt{Q(\theta_1,0)},
\sqrt{P(0,0)}+\sqrt{Q(0,0)} \right\}, & \texttt{if} & a_2= a_3.
\end{array} \right.
\end{array}
\end{equation}
\end{widetext}
 For $\varphi=\pi,2\pi$,  by (\ref{a1=0-varphi=pi-min}), there is
\begin{equation}\label{}
\begin{array}{ll}
  & \min\{\sqrt{P(\theta,\pi)}+\sqrt{Q(\theta,\pi)}\} \\
 = & \min\{\sqrt{P(\theta,2\pi)}+\sqrt{Q(\theta,2\pi)}\} \\
 = & \min\{\sqrt{P(\theta,0)}+\sqrt{Q(\theta,0)}\}.
\end{array}
\end{equation}.

Now we investigate the hyperplanes $\varphi=\frac{\pi}{2}$ and $\varphi=\frac{3\pi}{2}$. From
$\frac{\texttt{d}(\sqrt{P(\theta,\frac{\pi}{2})}+\sqrt{Q(\theta,\frac{\pi}{2})})}{\texttt{d}\theta}=0$, there is
$\cos\theta=0$, i.e.  $\theta=\frac{\pi}{2}$.  Similarly, by
$\frac{\texttt{d}(\sqrt{P(\theta,\frac{3\pi}{2})}+\sqrt{Q(\theta,\frac{3\pi}{2})})}{\texttt{d}\theta}=0$, we
also get $\theta=\frac{\pi}{2}$. It is direct from (\ref{q-p-relation}) and (\ref{p-q-relation}) that
\begin{equation}\label{}
 \begin{array}{ll}
   & \sqrt{P(\frac{\pi}{2},\frac{\pi}{2})}+\sqrt{Q(\frac{\pi}{2},\frac{\pi}{2})} \\
= &  \sqrt{P(\frac{\pi}{2},\frac{3\pi}{2})}+\sqrt{Q(\frac{\pi}{2},\frac{3\pi}{2})} \\
=& \sqrt{1-4a_2^2a_3^2+8a_1a_2a_3a_4-4a_0^2a_4^2-4a_1^2a_4^2}.
 \end{array}
\end{equation}

Therefore,
\begin{equation}\label{}
 \begin{array}{ll}
    & \min\{\sqrt{P(\theta,\varphi)}+\sqrt{Q(\theta,\varphi)}\} \\
  =  &  \min\{\sqrt{P(0,\varphi)}+\sqrt{Q(0,\varphi)}, \sqrt{P(\frac{\pi}{2},\frac{\pi}{2})}+\sqrt{Q(\frac{\pi}{2},\frac{\pi}{2})}, \\
    & ~~~~~~ \min\{\sqrt{P(\theta,0)}+\sqrt{Q(\theta,0)}\} \}. \\
 \end{array}
\end{equation}

\subsection{$\mu=\pi$ and $a_0a_1a_2a_3a_4\neq 0$}

In this section, we examine the quantum channel (\ref{quantumchannel}) with coefficients satisfying  $\mu=\pi$ and $a_0a_1a_2a_3a_4\neq 0$.

 Let us
start with the special case $\sin\theta=0$. In this case, there is
\begin{equation}\label{}
\begin{array}{ll}
& \sqrt{P(0,\varphi)}+\sqrt{Q(0,\varphi)}\\
 & =\sqrt{P(\pi,\varphi)}+\sqrt{Q(\pi,\varphi)} \\
  & =a_0^2+\sqrt{(1-a_0^2)^2-4(a_1a_4+a_2a_3)^2} \\
  & = a_0^2+\sqrt{[(a_2-a_3)^2+(a_1-a_4)^2][(a_2+a_3)^2+(a_1+a_4)^2]}, \\
\end{array}
\end{equation}
Evidently, $P(0,\varphi)=Q(\pi,\varphi)=a_0^2\neq 0$, $P(\pi,\varphi)=0$ if and only if $a_2=a_3$ and $a_1=a_4$,
and $Q(0,\varphi)=0$ if and only if $a_2=a_3$ and $a_1=a_4$. Next we suppose that $\sin\theta\neq 0$, i.e.
$\theta\in(0,\pi)$.

First, we look for the condition of $P(\theta,\varphi)=0$, $Q(\theta,\varphi)=0$. If $P(\theta,\varphi)=0$, then
\begin{widetext}\begin{equation}
\begin{array}{ll}
& \cos\varphi=z_1\\
 =  &  \frac{1}{2a_0a_1^2t}[t^2a_0^2a_1+(1-a_0^2)a_1-2a_4(a_2a_3+a_1a_4)  -2\sqrt{-t^2a_0^2a_1a_2a_3a_4+(a_2a_3+a_1a_4)(a_1a_3-a_2a_4)(a_1a_2-a_3a_4)} ], \\
& \cos\varphi=z_2\\
 =  &  \frac{1}{2a_0a_1^2t}[t^2a_0^2a_1+(1-a_0^2)a_1-2a_4(a_2a_3+a_1a_4)   +2\sqrt{-t^2a_0^2a_1a_2a_3a_4+(a_2a_3+a_1a_4)(a_1a_3-a_2a_4)(a_1a_2-a_3a_4)} ]. \\
\end{array}
\end{equation}\end{widetext}
Here $(a_1a_3-a_2a_4)(a_1a_2-a_3a_4)>0$, and $t\in
(0,\sqrt{\frac{(a_2a_3+a_1a_4)(a_1a_3-a_2a_4)(a_1a_2-a_3a_4)}{a_0^2a_1a_2a_3a_4}}]$, since
$(a_1a_3-a_2a_4)(a_1a_2-a_3a_4)\leq 0$ implies that $P(\theta,\varphi)\neq 0$. Directly,
$(a_1a_3-a_2a_4)(a_1a_2-a_3a_4)>0$ if and only if $a_2^2+a_3^2<\frac{a_2a_3(a_1^2+a_4^2)}{a_1a_4}$.

 Note that
\begin{equation*}
\begin{array}{ll}
  & (1-a_0^2)a_1-2a_4(a_2a_3+a_1a_4) \\
 = & a_1(a_1^2-a_4^2)+a_1(a_2^2+a_3^2)-2a_2a_3a_4 \\
 > & 0
\end{array}
\end{equation*}
in case of $a_1>a_4$,
\begin{equation*}
\begin{array}{ll}
  & (1-a_0^2)a_1-2a_4(a_2a_3+a_1a_4) \\
 = & a_1(a_1^2-a_4^2)+a_1(a_2^2+a_3^2)-2a_2a_3a_4 \\
 < & \frac{(a_1^2-a_4^2)(a_2a_3+a_1a_4)}{a_4} \\
 < & 0
\end{array}
\end{equation*}
in case of $a_1<a_4$ and $(a_1a_3-a_2a_4)(a_1a_2-a_3a_4)>0$, and
\begin{equation}\label{}
\begin{array}{ll}
   & [(1-a_0^2)a_1-2a_4(a_2a_3+a_1a_4)]^2\\
 & -\{2\sqrt{(a_2a_3+a_1a_4)(a_1a_3-a_2a_4)(a_1a_2-a_3a_4)}\}^2 \\
 = & a_1^2[(a_1-a_4)^2+(a_2-a_3)^2][(a_1+a_4)^2+(a_2+a_3)^2] \\
 > & 0. \\
\end{array}
\end{equation} It follows that
\begin{equation}\label{}
\begin{array}{ll}
   & (1-a_0^2)a_1-2a_4(a_2a_3+a_1a_4) \\
& +2\sqrt{(a_2a_3+a_1a_4)(a_1a_3-a_2a_4)(a_1a_2-a_3a_4)} \\
 = & (a_1-a_4)(a_1^2+2a_3^2+a_1a_4)+2a_3\sqrt{(a_1-a_4)^2(a_3^2+a_1a_4)} \\
 > & 0 \\
\end{array}
\end{equation}  and
\begin{equation}\label{}
\begin{array}{ll}
   & (1-a_0^2)a_1-2a_4(a_2a_3+a_1a_4)\\
& -2\sqrt{(a_2a_3+a_1a_4)(a_1a_3-a_2a_4)(a_1a_2-a_3a_4)} \\
 = & (a_1-a_4)(a_1^2+2a_3^2+a_1a_4)-2a_3\sqrt{(a_1-a_4)^2(a_3^2+a_1a_4)} \\
 > & 0 \\
\end{array}
\end{equation} in case of $a_1>a_4$;
\begin{equation}\label{}
\begin{array}{ll}
   & (1-a_0^2)a_1-2a_4(a_2a_3+a_1a_4)\\
&   +2\sqrt{(a_2a_3+a_1a_4)(a_1a_3-a_2a_4)(a_1a_2-a_3a_4)} \\
 = & (a_1-a_4)(a_1^2+2a_3^2+a_1a_4)+2a_3\sqrt{(a_1-a_4)^2(a_3^2+a_1a_4)} \\
 <& 0 \\
\end{array}
\end{equation}  and
\begin{equation}\label{}
\begin{array}{ll}
   & (1-a_0^2)a_1-2a_4(a_2a_3+a_1a_4)\\
 &  -2\sqrt{(a_2a_3+a_1a_4)(a_1a_3-a_2a_4)(a_1a_2-a_3a_4)} \\
 = & (a_1-a_4)(a_1^2+2a_3^2+a_1a_4)-2a_3\sqrt{(a_1-a_4)^2(a_3^2+a_1a_4)} \\
 < & 0 \\
\end{array}
\end{equation} in case of $a_1<a_4$. Thus,  if $a_1>a_4$, then both $z_1$ and $z_2$ go to $+\infty$ when $t\rightarrow 0$;
 if $a_1<a_4$, then both $z_1$ and $z_2$ go to $-\infty$ when $t\rightarrow 0$.

Now suppose that $a_1>a_4$. From $z_1=1$, there is
\begin{widetext}
\begin{equation}\label{}
\begin{array}{rc}
 [t^2a_0^2a_1-2ta_0a_1^2+(1-a_0^2)a_1-2a_4(a_2a_3+a_1a_4)]^2 & \\
  -[2\sqrt{-t^2a_0^2a_1a_2a_3a_4+(a_2a_3+a_1a_4)(a_1a_3-a_2a_4)(a_1a_2-a_3a_4)}]^2 & =  0,
\end{array}
\end{equation}
\end{widetext}
the solutions of which are
\begin{equation}\label{}
\begin{array}{ll}
 t=& t_{11}= \frac{a_1+a_4-\sqrt{-(a_2+a_3)^2}}{a_0}, \\
 t=& t_{12}=  \frac{a_1+a_4+\sqrt{-(a_2+a_3)^2}}{a_0}, \\
 t=& t_{13}= \frac{a_1-a_4-\sqrt{-(a_2-a_3)^2}}{a_0},  \\
 t=& t_{14}= \frac{a_1-a_4+\sqrt{-(a_2-a_3)^2}}{a_0}.  \\
\end{array}
\end{equation}
Here, $t_{11}$ and $t_{12}$ are imaginary numbers, and $t_{13}$ and $t_{14}$ are positive real numbers only if
$a_2=a_3$ and $a_1>a_4$. It is not difficult to check that $\theta_0$ is a root of $z_1=1$ and the minimum point
of $z_1$ in case of $a_1>a_4$ and $a_2=a_3$. Here $t=\cot\frac{\theta_0}{2}=\frac{a_1-a_4}{a_0}$.  $z_1\geq 1$
comes directly since $z=z_1$ is a continuous function of $t$, goes to $+\infty$ when $t\rightarrow 0$, and has
only one intersection point with straight line $z=1$, where the equality $z_1=1$ holds iff $a_2=a_3$,$a_1>a_4$
and $\cot\frac{\theta_0}{2}=\frac{a_1-a_4}{a_0}$. We can show that $z_2>1$ in case of $a_1>a_4$ in the same way.

Similarly, we can prove that if $a_1<a_4$, then $z_1<-1$, and  $z_2\leq -1$,   where $z_2=-1$ iff $a_2=a_3$,
$a_1<a_4$, and $t=\cot\frac{\overline{\theta}_0}{2}=\frac{-a_1+a_4}{a_0}$. Therefore, $P(\theta,\varphi)=0$ iff
$\varphi=0$ and $\theta=\theta_0$ in case of  $a_2=a_3$ and $a_1>a_4$,   or $\varphi=\pi$ and
$\theta=\overline{\theta}_0$ in case of $a_2=a_3$ and
 $a_1<a_4$. It can be derived that
\begin{equation}\label{}
\begin{array}{ll}
  & \left(\sqrt{P(\theta,\varphi)}+\sqrt{Q(\theta,\varphi)}\right)\big|_{P(\theta,\varphi)=0, ~ \theta\in(0,\pi)} \\
  = & \sqrt{Q(\theta_0,0)}|_{\{a_2=a_3, a_1>a_4\}} \\
 = & \sqrt{Q(\overline{\theta}_0,\pi)}|_{\{a_2=a_3, a_1<a_4\}} \\
= & \sqrt{(1-2a_4^2)^2+4a_3^2(-1+a_1^2+a_3^2-2a_1a_4+3a_4^2)}. \\
\end{array}
\end{equation}
Immediately, from (\ref{q-p-relation}) and (\ref{p-q-relation}), $Q(\theta,\varphi)=0$ iff $\varphi=\pi$ and
$\theta=\pi-\theta_0$ in case of $a_2=a_3$ and $a_1>a_4$, or $\varphi=0$ and $\theta=\pi-\overline{\theta}_0$ in
case of $a_2=a_3$ and $a_1<a_4$, and
\begin{equation}\label{}
\begin{array}{ll}
  & (\sqrt{P(\theta,\varphi)}+\sqrt{Q(\theta,\varphi)})|_{Q(\theta,\varphi)=0} \\
  = & (\sqrt{P(\theta,\varphi)}+\sqrt{Q(\theta,\varphi)})|_{P(\theta,\varphi)=0}. \\
 \end{array}
\end{equation}

From above, it can be seen that the quantum channel (\ref{quantumchannel}), the coefficients of which satisfy
$\mu=\pi$, $a_0a_1a_2a_3a_4\neq 0$ and  $a_2=a_3$, can be collapsed to a Bell state with probability
$p_1=p_2=\frac{a_0^2(1-a_0^2+3a_1^2-4a_1a_4+a_4^2)}{a_0^2+(a_1-a_4)^2}$ by Charlie's measurement in the basis
(\ref{measurebasis}) with $(\theta,\varphi)=(\theta_0,0)$ or $(\theta,\varphi)=(\pi-\theta_0,\pi)$  in case of
$a_1>a_4$,  or $(\theta,\varphi)=(\overline{\theta}_0,\pi)$ or $(\theta,\varphi)=(\pi-\overline{\theta}_0,0)$ in
case of $a_1<a_4$. In particular, this quantum channel can also be purified to an EPR pair with probability
$2a_1^2+2a_2^2$ via controller's measurement in the basis $|0\rangle,|1\rangle$ in case of $a_1=a_4$.

In the following, we suppose that $P(\theta,\varphi)Q(\theta,\varphi)\neq 0$. In order to obtain the minimum of
$\sqrt{P(\theta,\varphi)}+\sqrt{Q(\theta,\varphi)}$, we need to find the points such that equations (\ref{diff-theta-varphi}) and
(\ref{diff-varphi}) hold.    From (\ref{diff-theta-varphi}), we get $ \sin\varphi=0$, or
\begin{equation}\label{mu=pi-d-theta-varphi}
\begin{array}{ll}
   &ya_0[2 a_1 a_2^2 a_3^2 - a_2 a_3 a_4(-2 a_0^2 - 4 a_1^2 - 2 a_4^2 + 1)\\
&  -  2 a_1 a_4^2(a_2^2 + a_3^2)] \\
   & -2\cos\varphi(a_2a_3+a_1a_4)(a_1a_3-a_2a_4)(a_1a_2-a_3a_4)\\
 = & 0. \\
\end{array}
\end{equation}
By (\ref{diff-varphi}), we have $ \sin\varphi=0$, or {\small   \begin{equation}\label{mu=pi-d-varphi}
\begin{array}{ll}
   & 4y^3a_0^3(a_2a_3+a_1a_4)(a_1a_3-a_2a_4)(a_1a_2-a_3a_4) \\
   & -8y^2a_0^2a_1\cos\varphi(a_2a_3+a_1a_4)(a_1a_3-a_2a_4)(a_1a_2-a_3a_4) \\
   & +ya_0[(a_1-2a_2a_3a_4-2a_1a_4^2)(2{a_1}a_2^2a_3^2 - {a_2}{a_3}{a_4} \\
   & + 4a_1^2{a_2}{a_3}{a_4}- 2{a_1}a_2^2a_4^2 - 2{a_1}a_3^2a_4^2 + 2{a_2}{a_3}a_4^3) \\
 & +4a_1^2\cos^2\varphi(a_2a_3+a_1a_4)(a_1a_3-a_2a_4)(a_1a_2-a_3a_4)]\\
  & +2\cos\varphi(-a_1+2a_2a_3a_4+2a_1a_4^2)[a_2^2 a_3^2(a_1^2 + a_4^2) \\
 &  - a_1 a_2  a_3a_4(-2 a_1^2 - 2 a_4^2 + 1) - a_1^2 (a_2^2 + a_3^2) a_4^2] \\
 = & 0.
\end{array}
\end{equation}}

If $(a_1a_3-a_2a_4)(a_1a_2-a_3a_4)=0$, then (\ref{mu=pi-d-theta-varphi}) and (\ref{mu=pi-d-varphi}) become
\begin{equation}\label{}
  y=0,
\end{equation}
and
 \begin{equation}\label{}
y[a_4^2-(1-2a_0^2)a_1^2]-2a_0a_1\cos\varphi(a_1^2+a_4^2)=0,
\end{equation}
respectively. However, if  $(a_1a_3-a_2a_4)(a_1a_2-a_3a_4)\neq 0$, then  (\ref{mu=pi-d-theta-varphi}) and (\ref{mu=pi-d-varphi}) become
{\footnotesize   \begin{equation}\label{}
\begin{array}{cl}
   &  \cos\varphi \\
 = & \frac{ya_0\{-2{a_1}a_4^2( a_3^2 + a_2^2) +
  {a_2}{a_3}[ 2{a_1}{a_2}{a_3} + {a_4} +
     2(a_1^2 -a_2^2 -a_3^2 ){a_4} ]\}}{2(a_2a_3+a_1a_4)(a_1a_3-a_2a_4)(a_1a_2-a_3a_4)} \\
\end{array}
\end{equation}}
and
\begin{equation}\label{}
\frac{y(1+y^2)a_0^3a_2^2a_3^2a_4^2(2a_2a_3a_4-a_1(1-2a_4^2))^2}{(a_2a_3+a_1a_4)(a_1a_3-a_2a_4)(a_1a_2-a_3a_4)}=0,
\end{equation}
respectively. That is, $y=0$, $\cos\varphi=0$.

For getting the minimum point of $\sqrt{P(\theta,\varphi)}+\sqrt{Q(\theta,\varphi)}$, it is enough for us to
consider the hyperplanes $\sin\varphi=0$, and $\cos\varphi=0$. For hyperplane $\sin\varphi=0$, we need only to
consider the case $\varphi=0$ by (\ref{q-p-relation}), (\ref{p-q-relation}), and (\ref{0-pi-2pi}). For
hyperplane $\cos\varphi=0$, we need only to consider the case $\varphi=\frac{\pi}{2}$ by (\ref{q-p-relation})
and (\ref{p-q-relation}).

Now, let us consider the case $\varphi=0$. In this case (\ref{diff-theta}) becomes
\begin{equation}\label{mu=pi-varphi=0-d-theta}
 \begin{array}{ll}
  & [-2a_1a_2a_3+a_4-2a_1^2a_4+2ya_0(a_2a_3+a_1a_4)]\\
& \times[4\,y^2\,a_0^2( {a_1}\,{a_3} - {a_2}\,{a_4}) ( {a_1}\,{a_2} - {a_3}\,{a_4}) \\
  & + 2\,y\,{a_0}( 2\,{a_1}\,{a_2}\,{a_3} - 4\,a_1^3\,{a_2}\,{a_3} - 2\,{a_1}\,a_2^3\,{a_3}\\
 &   - 2\,{a_1}\,{a_2}\,a_3^3- a_2^2\,{a_4}+ 4\,a_1^2\,a_2^2\,{a_4} + 2\,a_2^4\,{a_4} -
     a_3^2\,{a_4}  \\
 &  + 4\,a_1^2\,a_3^2\,{a_4} + 2\,a_3^4\,{a_4}- 8\,{a_1}\,{a_2}\,{a_3}\,a_4^2 +
     2\,a_2^2\,a_4^3 + 2\,a_3^2a_4^3 ) \\
 &  +({a_2}-2\,a_1^2\,{a_2}- 2\,a_2^3+ 2{a_1}{a_3}{a_4}- 2{a_2}\,a_4^2 ) \\
 & \times({a_3}- 2\,a_1^2{a_3}- 2\,a_3^3 + 2\,{a_1}\,{a_2}{a_4} - 2{a_3}a_4^2)] \\
 =&  0. \\
 \end{array}
\end{equation}
If $(a_1a_3-a_2a_4)(a_1a_2-a_3a_4)\neq 0$, then
 \begin{equation}\label{}
\begin{array}{ll}
  y= & \cot\theta_1= \frac{2\,{a_1}\,{a_2}\,{a_3} - {a_4} + 2a_1^2\,{a_4}}{2\,{a_0}\,\left( {a_2}\,{a_3} + {a_1}\,{a_4} \right) },  \\
  y= & \cot\theta_2=\frac{-{a_2} + 2a_1^2\,{a_2} + 2a_2^3 - 2\,{a_1}\,{a_3}\,{a_4} + 2\,{a_2}a_4^2}
  {2\,{a_0}\,\left( {a_1}\,{a_2} - {a_3}\,{a_4} \right) }, \\
  y= & \cot\theta_3= \frac{-{a_3} + 2a_1^2\,{a_3} + 2a_3^3 - 2\,{a_1}\,{a_2}\,{a_4} + 2\,{a_3}a_4^2}
  {2\,{a_0}\,\left( {a_1}\,{a_3} - {a_2}\,{a_4} \right) }, \\
\end{array}
\end{equation}
where $\theta_1$ is a root of equation (\ref{differencial-theta}), while $\theta_2$ is a root of equation
(\ref{differencial-theta}) only in case of $a_2<a_3$, and $\theta_3$ is a root of equation
(\ref{differencial-theta}) only in case of  $a_2>a_3$.
 If $( {a_1}\,{a_2} - {a_3}\,{a_4})= 0$, and $a_1\neq a_4$, then $a_2\neq a_3$, and
 \begin{equation}\label{}
\begin{array}{rl}
y & =\cot\theta_4 =\frac{-a_1+2a_1^3+2a_1a_3^2}{2a_0(a_1^2+a_3^2)}, \\
 y & =\cot\theta_5=\frac{-a_1+2a_1^3+2a_1a_3^2}{2a_0(a_1^2-a_4^2)},\\
\end{array}
\end{equation}
where  $\theta_4$ is a root of equation (\ref{differencial-theta}), while $\theta_5$ is a root of equation
(\ref{differencial-theta}) only if $a_2>a_3$.
 If $( {a_1}\,{a_3} - {a_2}\,{a_4})= 0$, and $a_1\neq a_4$, then $a_2\neq a_3$, and
 \begin{equation}\label{}
\begin{array}{rl}
y & =\cot\theta_6=\frac{-a_1+2a_1^3+2a_1a_2^2}{2a_0(a_1^2+a_2^2)}, \\
 y & =\cot\theta_7=\frac{-a_1+2a_1^3+2a_1a_2^2}{2a_0(a_1^2-a_4^2)},\\
\end{array}
\end{equation}
where $\theta_6$ is a root of equation (\ref{differencial-theta}), while $\theta_7$ is a root only if $a_2<a_3$.
 If $a_1=a_4$ and $a_2=a_3$, then
\begin{equation}\label{}
y=\cot\theta_8=-\frac{a_0a_1}{2(a_1^2+a_2^2)},
\end{equation}
where $\theta_8$ is  a root of equation (\ref{differencial-theta}).

Therefore,
\begin{equation}\label{mu=pi-varphi=0-min}
\begin{array}{ll}
  & \min\big\{ \sqrt{P(\theta,0)}+\sqrt{Q(\theta,0)}\big \} \\
= & \left\{  \begin{array}{lll}
 \min\{\left(\sqrt{P(\theta,0)}+\sqrt{Q(\theta,0)}\right)|_{a_2>a_3} \}, & \texttt{if} & a_2>a_3,\\
 \min\{\left(\sqrt{P(\theta,0)}+\sqrt{Q(\theta,0)}\right)|_{a_2<a_3} \}, & \texttt{if} & a_2<a_3,\\
  \min\{\left(\sqrt{P(\theta,0)}+\sqrt{Q(\theta,0)}\right)|_{a_2=a_3} \}, & \texttt{if} & a_2=a_3.\\
\end{array} \right.
\end{array}
\end{equation}
Here,
\begin{widetext}
\begin{equation}
\begin{array}{ll}
  & \min\{ \left(\sqrt{P(\theta,0)}+\sqrt{Q(\theta,0)}\right)|_{a_2>a_3} ~ \} \\
 = & \left\{\begin{array}{lll}
   \min\{ \sqrt{P(\theta_1,0)}+\sqrt{Q(\theta_1,0)}, \sqrt{P(\theta_3,0)}+\sqrt{Q(\theta_3,0)}, \sqrt{P(0,0)}+\sqrt{Q(0,0)} \}, & \texttt{if} & (a_1a_3-a_2a_4)(a_1a_2-a_3a_4)\neq 0, \\
   \min\{ \sqrt{P(\theta_4,0)}+\sqrt{Q(\theta_4,0)}, \sqrt{P(\theta_5,0)}+\sqrt{Q(\theta_5,0)}, \sqrt{P(0,0)}+\sqrt{Q(0,0)} \}, & \texttt{if} & a_1a_2-a_3a_4=0 ~ \texttt{and} ~ a_1\neq a_4,\\
 \min\{ \sqrt{P(\theta_6,0)}+\sqrt{Q(\theta_6,0)}, \sqrt{P(0,0)}+\sqrt{Q(0,0)} \},  & \texttt{if} & a_1a_3-a_2a_4=0 ~ \texttt{and} ~ a_1\neq a_4;\\
 \end{array}\right.
\end{array}
\end{equation}
\begin{equation}
\begin{array}{ll}
  & \min\{ \left(\sqrt{P(\theta,0)}+\sqrt{Q(\theta,0)}\right)|_{a_2<a_3} ~ \} \\
 = & \left\{\begin{array}{lll}
   \min\{ \sqrt{P(\theta_1,0)}+\sqrt{Q(\theta_1,0)}, \sqrt{P(\theta_2,0)}+\sqrt{Q(\theta_2,0)}, \sqrt{P(0,0)}+\sqrt{Q(0,0)}\}, & \texttt{if} & (a_1a_3-a_2a_4)(a_1a_2-a_3a_4)\neq 0, \\
 \min\{   \sqrt{P(\theta_4,0)}+\sqrt{Q(\theta_4,0)}, \sqrt{P(0,0)}+\sqrt{Q(0,0)} \},  & \texttt{if} & a_1a_2-a_3a_4=0 ~ \texttt{and} ~ a_1\neq a_4,\\
    \min\{ \sqrt{P(\theta_6,0)}+\sqrt{Q(\theta_6,0)}, \sqrt{P(\theta_7,0)}+\sqrt{Q(\theta_7,0)}, \sqrt{P(0,0)}+\sqrt{Q(0,0)} \}, & \texttt{if} & a_1a_3-a_2a_4=0 ~ \texttt{and} ~ a_1\neq a_4;\\
  \end{array}\right.
\end{array}
\end{equation}
\begin{equation}
\begin{array}{ll}
  & \min\{ \left(\sqrt{P(\theta,0)}+\sqrt{Q(\theta,0)}\right)|_{a_2=a_3} ~ \} \\
 = & \left\{\begin{array}{lll}
 \min\{  \sqrt{P(\theta_1,0)}+\sqrt{Q(\theta_1,0)}, \sqrt{Q(\theta_0, 0)}|_{a_2=a_3, a_1>a_4}, \sqrt{P(0,0)}+\sqrt{Q(0,0)}  \}, & \texttt{if} & a_1> a_4, \\
 \min\{  \sqrt{P(\theta_1,0)}+\sqrt{Q(\theta_1,0)}, \sqrt{Q(\overline{\theta}_0, \pi)}|_{a_2=a_3, a_1<a_4}, \sqrt{P(0,0)}+\sqrt{Q(0,0)}  \}, & \texttt{if} & a_1< a_4, \\
 \min\{   \sqrt{P(\theta_8,0)}+\sqrt{Q(\theta_8,0)}, a_0^2 \}, & \texttt{if} &  a_1= a_4.\\
 \end{array}\right.
\end{array}
\end{equation}
\end{widetext}

By (\ref{q-p-relation}), (\ref{p-q-relation}) and (\ref{0-pi-2pi}), we have
\begin{widetext}
\begin{equation}\label{}
\begin{array}{ll}
 \min\{ \sqrt{P(\theta,\pi)}+\sqrt{Q(\theta,\pi)} \}= & \min\{ \sqrt{P(\theta,0)}+\sqrt{Q(\theta,0)} \}=\min\{ \sqrt{P(\theta,2\pi)}+\sqrt{Q(\theta,2\pi)} \} \\
\end{array}
\end{equation}
\end{widetext}

Now let us examine  the case $\varphi=\frac{\pi}{2}$. In this case, (\ref{diff-theta}) goes to
\begin{eqnarray}
&4ya_0^4[1-2a_0^2-4(a_2a_3+a_1a_4)^2][a_2^2a_3^2 & \nonumber\\
 &+2a_1a_2a_3a_4-a_4^2(a_2^2+a_3^2)](1+y^2)^{-\frac{3}{2}}=& 0,
\end{eqnarray}
which implies that $\theta=\frac{\pi}{2}$. It is shown that $\theta=\frac{\pi}{2}$ is the only possible extreme
point of both $\sqrt{P(\theta,\frac{\pi}{2})}+\sqrt{Q(\theta,\frac{\pi}{2})}$ and
$\sqrt{P(\theta,\frac{3\pi}{2})}+\sqrt{Q(\theta,\frac{3\pi}{2})}$, and
\begin{equation}\label{}
\begin{array}{ll}
 & \sqrt{P(\frac{\pi}{2},\frac{\pi}{2})}+\sqrt{Q(\frac{\pi}{2},\frac{\pi}{2})}  \\
= & \sqrt{P(\frac{\pi}{2},\frac{3\pi}{2})}+\sqrt{Q(\frac{\pi}{2},\frac{3\pi}{2})}  \\
 = &\sqrt{1-4(a_2a_3+a_1a_4)^2-4a_0^2a_4^2}. \\
\end{array}
\end{equation}

To sum up,
\begin{equation}\label{mu=0-min}
\begin{array}{ll}
  &\min\{ \sqrt{P(\theta,\varphi)}+\sqrt{Q(\theta,\varphi)}\} \\
  = & \min\Big\{ \sqrt{P(0,\varphi)}+\sqrt{Q(0,\varphi)}, \sqrt{P(\frac{\pi}{2},\frac{\pi}{2})}+\sqrt{Q(\frac{\pi}{2},\frac{\pi}{2})}, \\
  & ~~~~~~~ \min\{\sqrt{P(\theta,0)}+\sqrt{Q(\theta,0)}\}, \\
 & ~~~~~~~  (\sqrt{P(\theta,\varphi)}+\sqrt{Q(\theta,\varphi)})|_{P(\theta,\varphi)=0}\Big\}. \\
\end{array}
\end{equation}

\section{Conclusion}

In brief, we give the analytic expression of the localizable entanglement (LE), the maximum of probability of
successfully controlled teleporting an unknown qubit state (\ref{unknownstate}) via every three-qubit state
(\ref{quantumchannel}) satisfying $a_1a_2a_3a_4\sin\mu =0$ and investigate how to achieve it (that is, Charlie
finds optimal measurement basis, performs optimal measurements, and communicates the results).

The authors thank Prof. J. I. Cirac for his fruitful discussions and for his hospitality during their  stay at
Max-Planck-Institut f\"{u}r Quantenoptik. This work was supported by  the National Natural Science Foundation of
China under Grant No: 10671054, Hebei Natural
 Science Foundation of China
under Grant Nos: A2005000140, 07M006, and the Key Project of Science and Technology Research of Education Ministry of
 China under Grant No:
207011.

\end{document}